\documentclass[prx,aps,groupedaddress,twocolumn,superscriptaddress,longbibliography]{revtex4-2}
\usepackage{graphicx}
\usepackage{amsmath}
\usepackage{amssymb}

\usepackage[dvipsnames]{xcolor}
\usepackage{changepage}
\usepackage{tcolorbox}
\usepackage{pifont}
\usepackage{enumitem}

\newlist{todolist}{itemize}{2}
\setlist[todolist]{label=$\square$}

\newcommand{\gsystem}[3]{#1\!\!\times\!\!#2\!\!\times\!\!#3}

\newcommand{\equref}[1]{Eq.~(\ref{eq:#1})}
\newcommand{\figref}[1]{Fig.~\ref{fig:#1}}
\newcommand{\appref}[1]{App.~\ref{app:#1}}
\newcommand{\sectref}[1]{Sect.~\ref{sect:#1}}
\newcommand{\tabref}[1]{Tab.~\ref{tab:#1}}

\usepackage[colorlinks=true,linkcolor=blue]{hyperref}
\hypersetup{
	colorlinks,
	linkcolor={blue!75!black},
	citecolor={blue!75!black},
	urlcolor={blue!75!black},
}

\usepackage{tikz}
\usetikzlibrary{decorations.markings}

\newcommand{\bigbra}[1]{\big\langle#1\big|}
\newcommand{\bigket}[1]{\big|#1\big\rangle}

\newcommand{\ket}[1]{|\,#1\,\rangle}
\newcommand{\overlap}[2]{\langle\,#1\,|\,#2\,\rangle}
\newcommand{\cre}[1]{c^\dagger_{#1}}
\newcommand{\ani}[1]{c_{#1}}
\newcommand{\gs}[1]{\ket{\mathrm{GS_{\lambda=#1}}}}
\renewcommand{\i}{\mathrm{i}}
\newcommand{\e}{\mathrm{e}}

\newcommand{\flipA}{
	\begin{tikzpicture}[baseline=-0.25em,scale=1.15]
	\def\x{0.0};
	\def\y{-0.2};
	\def\L{0.4};
	\begin{scope}[decoration={
	    markings,
	    mark=at position 0.6 with {\arrow{<}}}
	    ]
	    \draw[postaction={decorate}] (\x,\y)--(\x+\L,\y);
	    \draw[postaction={decorate}] (\x+\L,\y)--(\x+\L,\y+\L);
	    \draw[postaction={decorate}] (\x+\L,\y+\L)--(\x,\y+\L);
	    \draw[postaction={decorate}] (\x,\y+\L)--(\x,\y);
	\end{scope}
	\end{tikzpicture}
}

\newcommand{\flipB}{
	\begin{tikzpicture}[baseline=-0.25em,scale=1.15]
	\def\x{0.0};
	\def\y{-0.2};
	\def\L{0.4};
	\begin{scope}[decoration={
	    markings,
	    mark=at position 0.6 with {\arrow{>}}}
	    ]
	    \draw[postaction={decorate}] (\x,\y)--(\x+\L,\y);
	    \draw[postaction={decorate}] (\x+\L,\y)--(\x+\L,\y+\L);
	    \draw[postaction={decorate}] (\x+\L,\y+\L)--(\x,\y+\L);
	    \draw[postaction={decorate}] (\x,\y+\L)--(\x,\y);
	\end{scope}
	\end{tikzpicture}
}

\newcommand{\flipAParticle}{
	\begin{tikzpicture}[baseline=-0.25em,scale=1.15]
	\def\x{0.0};
	\def\y{-0.2};
	\def\L{0.4};
	\draw (\x,\y) -- (\x+\L,\y) -- (\x+\L,\y+\L) -- (\x,\y+\L) -- (\x,\y);

	\def\r{0.15em};
	\draw[black,fill=black] (\x,\y+\L/2) circle (\r);
	\draw[black,fill=black] (\x+\L/2,\y+\L) circle (\r);
	\end{tikzpicture}
}

\newcommand{\flipBParticle}{
	\begin{tikzpicture}[baseline=-0.25em,scale=1.15]
	\def\x{0.0};
	\def\y{-0.2};
	\def\L{0.4};
	\draw (\x,\y) -- (\x+\L,\y) -- (\x+\L,\y+\L) -- (\x,\y+\L) -- (\x,\y);

	\def\r{0.15em};
	\draw[black,fill=black] (\x+\L/2,\y) circle (\r);
	\draw[black,fill=black] (\x+\L,\y+\L/2) circle (\r);
	\end{tikzpicture}
}

\begin{document}
\title{Exploring Bosonic and Fermionic Link Models on $(3+1)-$d tubes}

\author{Debasish Banerjee}
\affiliation{Theory Division, Saha Institute of Nuclear Physics, HBNI, 1/AF Bidhannagar, Kolkata 700064}

\author{Emilie Huffman}
\affiliation{Perimeter Institute for Theoretical Physics, 31 Caroline Street North, Waterloo, Ontario, Canada N2L 2Y5}

\author{Lukas Rammelm\"{u}ller}
\affiliation{Arnold Sommerfeld Center for Theoretical Physics (ASC), University of Munich, Theresienstr. 37, 80333 M\"unchen, Germany}
\affiliation{Munich Center for Quantum Science and Technology (MCQST), Schellingstr. 4, 80799 M\"unchen, Germany}

\date{\today}

\begin{abstract}
  Quantum link models (QLMs) have attracted a lot of attention in recent times as a
  generalization of Wilson's lattice gauge theories (LGT), and are particularly
  suitable for realization on quantum simulators and computers. These models are known
  to host new phases of matter and act as a bridge between particle and condensed matter
  physics. In this article, we study the Abelian $U(1)$ lattice gauge theory in $(3+1)$-d
  tubes using large-scale exact diagonalization (ED). We are then able to motivate the
  phase diagram of the model with finite size scaling techniques (FSS), and in particular
  propose the existence of a Coulomb phase. Furthermore, we introduce the first models
  involving \emph{fermionic quantum links}, which generalize the gauge degrees of freedom
  to be of fermionic nature. We prove that while the spectra remain identical between 
  the bosonic and the fermionic versions of the $U(1)$-symmetric quantum link models in 
  $(2+1)$-d, they are different in $(3+1)$-d. We discuss the prospects of realizing the 
  magnetic field interactions as correlated hopping in quantum simulator experiments.
\end{abstract}

 \maketitle
 \tableofcontents

\begin{figure}
    \centering
    \includegraphics[width=0.95\columnwidth]{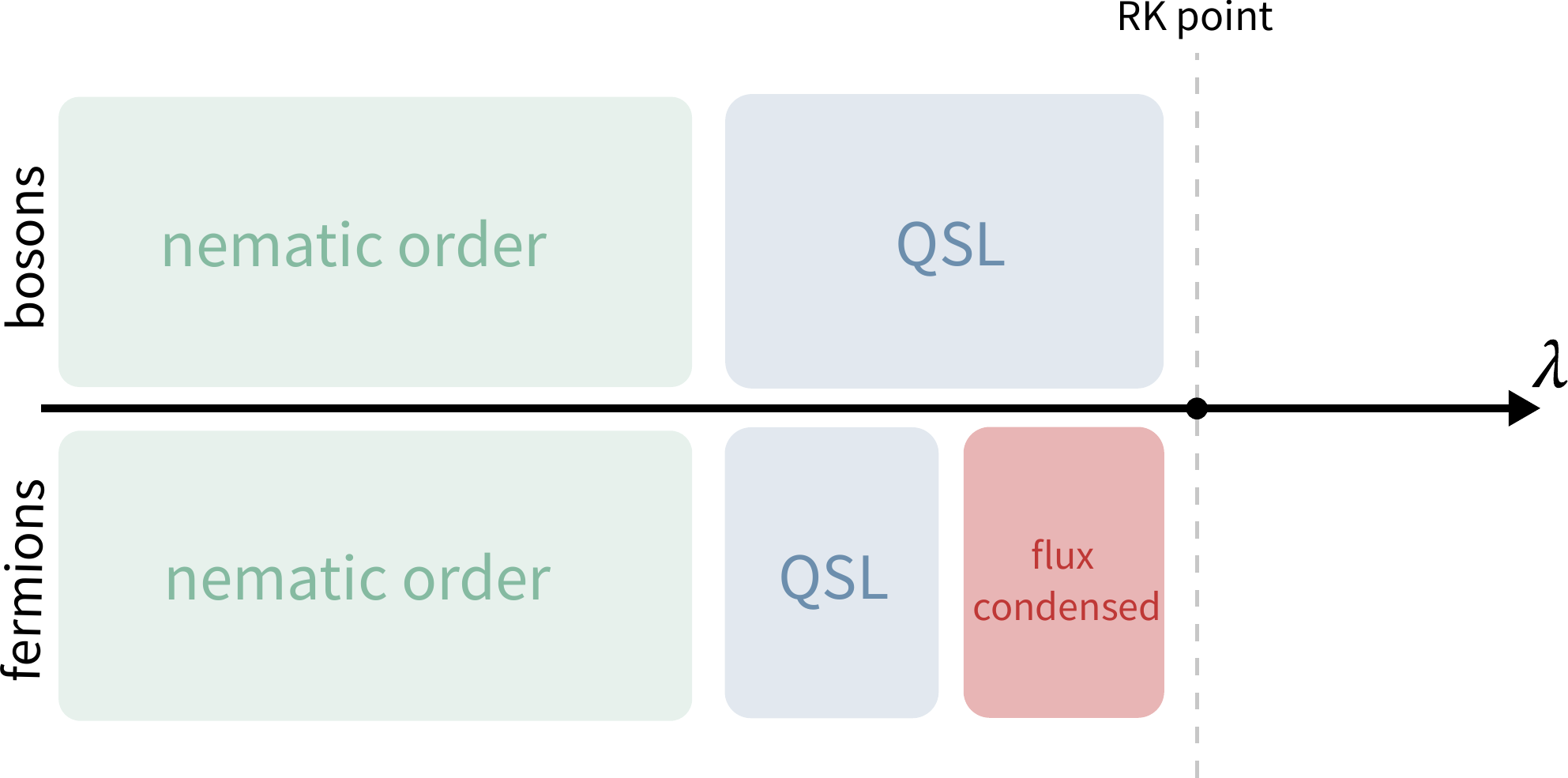}
    \caption{{\bf A schematic phase diagram of the $U(1)$ quantum link model} with the 
    spin and fermionic representation as a function of the coupling $\lambda$. For large 
    negative $\lambda$, there is a nematically ordered phase  which spontaneously breaks 
    the lattice (rotation) symmetries. For small values of $\lambda$, the broken symmetry 
    is restored. For $\lambda \rightarrow 1$, the winding fluxes can be excited easily and 
    additionally for the fermionic model there is a narrow region where they become the 
    ground state, before hitting the Rokshar-Kivelson point at $\lambda=1$.}
    \label{fig:pd_sketch}
\end{figure}

\section{Introduction}
  Gauge theories have a formidable legacy in the description of
  naturally occurring matter. Examples of their diverse applications include ab-initio
  descriptions of the strong interaction phenomenology in particle physics, which use
  quantum chromodynamics (QCD) as the starting point, and descriptions of superconductivity
  in condensed matter physics, which use $U(1)$ gauge fields to bind electrons. Even the 
  Kitaev model, which is used to introduce topological quantum computation, is a 
  $\mathbb{Z}_2$ lattice gauge theory. Naturally, many of these gauge theories need to 
  couple the fundamental degrees of freedom very strongly, which in turn renders 
  weak-coupling perturbation theory useless. Wilson \cite{Wilson1974} pioneered the 
  technique of discretizing the gauge theories non-perturbatively on a space-time lattice 
  and the use of Markov chain Monte Carlo methods to sample the resulting path integral. 
  This approach has been developed to a high degree of sophistication, where many aspects of 
  particle physics and condensed matter phenomenology can be directly studied ab-initio 
  using Monte Carlo simulations \cite{Wilczek1982}.

  While in the Wilsonian approach, one discretizes the gauge field action via the parallel
  transporters which live on the links of a lattice, it is also possible to approach the 
  problem from a Hamiltonian perspective. The latter approach \cite{Kogut1975} is 
  particularly useful when one wants to address gauge theories using the novel tools of 
  quantum simulators and quantum computers. Quantum computing is a rapidly developing
  computing paradigm using the notions of quantum entanglement, and can in principle highly 
  outperform classical computing paradigms (such as Markov Chain Monte Carlo) in certain 
  parameter regimes \cite{Banuls2020}. Such regimes occur in strongly correlated systems 
  for unitary evolution of the system in real time, at finite density, or at background 
  electric and magnetic fields.

  In the Hamiltonian formulation of Wilson's lattice gauge theories for compact $U(1)$ or 
  $SU(2)$ gauge groups, one has to deal with a locally infinite-dimensional Hilbert space 
  even for single link degrees of freedom. This makes it tricky to use this formulation
  for quantum computation, which naturally has a finite number of available states. Imposing 
  a naive cut-off on the number of allowed states risks breaking gauge invariance. Remarkably, 
  it is possible to define gauge theories that have finite-dimensional Hilbert spaces, and 
  yet still possess these continuous gauge symmetries, by judiciously using non-unitary link 
  operators. Quantum link models (QLMs), as they are called, have been theoretically developed 
  to possess both Abelian and non-Abelian local symmetries
  \cite{Horn1981,Orland1990, Chandrasekharan1997}, including QCD \cite{Brower1999}.  
  Qubit-regularized quantum field theories (QFTs) which generalize the construction of QFTs 
  using discrete degrees of freedom are being actively investigated \cite{Singh2019, Liu2021}.
  Such formulations have also been used in condensed matter physics in the context of
  superconductivity \cite{Kivelson1987, Moessner2011} and frustrated magnetism
  \cite{Moessner2003}. Only recently have the connections between the corresponding 
  microscopic theories been fully appreciated, and exploited to better understand the 
  underlying physics of the systems. The fact that they have a
  finite-dimensional Hilbert space of the gauge degrees of freedom, and yet still possess
  the same local symmetry as the Wilson-type models, makes them attractive candidates as
  models to be implemented in quantum computers, or quantum simulators.

  While it is established that these QLMs can be set up to have the same continuous gauge
  symmetries as the Wilson-type theories, there remain many open questions as to the nature
  of the phases that this family of gauge theories can host. Since they are generalized
  lattice gauge theories, they certainly give rise to novel phases which cannot be realized
  on Wilson-type theories. As a classic example, $U(1)$  QLMs in $(2+1)-$d give rise to
  phases where electric flux tubes joining static charges are fractionalized in units of
  $1/2$ or even $1/4$ \cite{Banerjee2013,Banerjee2014}. However, whether QLMs in higher
  dimensions can support deconfined Coulomb phases like continuum gauge fields is still
  an open question. A resolution of this question would certainly boost the importance
  of QLMs for consideration in quantum simulator experiments.

  Interestingly, the same question is also of prime importance in condensed matter
  physics, where existence of the Coulomb phase is a key ingredient to postulate the
  existence of quantum spin liquids, a phase of matter which does not break any
  internal or lattice symmetries and has fractionalized excitations. Previous work
  has already provided indications that this might indeed be the case
  \cite{Hermele2004,Sikora2009,Sikora2011}. In this article, using large-scale exact
  diagonalization on the $U(1)$ QLM on the cubic lattice and techniques of finite
  size scaling, we provide evidence of a region in the parameter space where the ground 
  state does not  break any symmetries, lattice or internal, and the system is gapped. 
  Since we are severely restricted to small lattice sizes, our results should also 
  encourage the development of novel algorithms to address the system on large lattices, 
  or perhaps quantum simulator experiments. Should the different computational methods be 
  able to establish the existence of a Coulomb phase in these models, it would be of 
  fundamental importance in the context of quantum  field theories as well. We would 
  thus have demonstrated an intriguing way to generate a massless gauge boson from a 
  microscopic theory with a finite-dimensional Hilbert space.

  As an important conceptual development, we also extend the same ideas which inspired the
  quantum link formulation to introduce a novel kind of quantum link model, where
  the gauge link operators are represented by fermionic creation and annihilation
  operators. We emphasize that this construction is distinct
  from the rishon representation \cite{Brower1999, Banerjee2013a}, where the quantum 
  link operators are represented as fermionic bilinear operators, such that all commutation 
  relations are preserved. We establish that the gauge invariance in the model is a 
  consequence of the special type of correlated sub-dimensional hopping of the fermionic 
  particles living on the links, and thus connected to similar ideas in the models of 
  fractons. Using  geometric constructions, we show that in $(2+1)-$d, the spectra of 
  the fermions and quantum spins $S = \frac{1}{2}$ are identical, while in $(3+1)-$d 
  they differ due to the Pauli exclusion principle. Using ED studies, we also offer a 
  first glimpse into the phase diagram of the fermionic model in $(3+1)-$d. One expects 
  that any reasonably efficient quantum Monte Carlo method that can be made to work for 
  the spin model would suffer a severe sign problem for the fermionic version
  \cite{Troyer2005}, so we outline the possibility of realizing this Hamiltonian on an 
  analog quantum simulator platform.

\section{Models and Symmetries}
We begin our presentation by describing the microscopic models and the symmetries
of the system. We also motivate how these models can be applied to naturally occurring
phenomena in particle and condensed matter physics.

\subsection{Bosonic Quantum Link Model}
 We first introduce the conventional bosonic version of an $U(1)$ Abelian QLM. While these 
 models can be studied on any lattice on which loops can be defined, we consider square
 and cubic lattices for concreteness. The operators of the gauge theory are defined on the
 links joining two adjacent lattice sites. The Hamiltonian of the link model is
 \begin{equation} \label{u1H}
 \begin{aligned}
   H = \frac{g^2}{2} \sum_{x,\mu} E^2_{x,\mu} &- J \sum_{\square} (U_\square + U^\dagger_\square) \\
   & + \lambda \sum_{\square} (U_\square + U^\dagger_\square)^2
 \end{aligned}
 \end{equation}
 where $E_{x,\mu}$ is the electric field operator defined on the link joining the sites $x$ 
 and $x+\hat{\mu}$. The first term is the electric field energy, the second term expressed 
 via plaquettes is the magnetic field energy, and the third term is the Rokhsar-Kivelson (RK)
 term. The plaquette operator, $U_\square$, is defined via the parallel transport operator
 $U_{x,\mu}$ as:
  \begin{equation} \label{plaq}
   U_\square = U_{x,\mu} U_{x+\hat{\mu},\nu} U^\dagger_{x+\hat{\nu},\mu} U^\dagger_{x,\nu}.
  \end{equation}
  Each link has three operators $U_{x,\mu}$, $U^\dagger_{x,\mu}$, and $E_{x,\mu}$ which can 
  be realized by the generators of an $SU(2)$ algebra. The operators satisfy the following
  commutation relations:
  \begin{align} \label{comm1}
   [E_{x,\mu},U_{y,\nu}] &= U_{x,\mu} \delta_{x,y} \delta_{\mu,\nu}\nonumber \\
   [E_{x,\mu},U^\dagger_{y,\nu}] &= -U^\dagger_{x,\mu} \delta_{x,y} \delta_{\mu,\nu} \\
   [U_{x,\mu},U^\dagger_{y,\nu}] &= 2 E_{x,\mu} \delta_{x,y} \delta_{\mu,\nu} \nonumber.
  \end{align}
The Hamiltonian has a local $U(1)$ invariance generated by the lattice Gauss law operator
\begin{equation} \label{gauss1}
   G_x = \sum_{\mu} (E_{x,\mu} - E_{x-\hat{\mu},\mu}),
\end{equation}
with the local commutation relations
\begin{equation}
   [G_x,H] = 0,\quad \text{for all}~~x.
\end{equation}
This necessitates the specification of additional conditions to define the superselection
sector of the Hilbert space by specifying the local charges. In the context of particle
physics, it is usual to choose a vacuum which does not have any charges. Mathematically, this
is expressed as:  $G_x \ket{\psi}= 0$, where $\ket{\psi}$ is a physical state of the theory. 
It is, of course, possible to choose various other superselection sectors by specifying
different charges on different sites. An example is the quantum dimer model, a model to
describe the non-N\'eel phases of quantum anti-ferromagnets relevant to high-T$_c$
superconductivity. This model works with a different superselection sector, mathematically
represented as $G_x \ket{\chi}= (-1)^x\ket{\chi}$, where $(-1)^x$ is the site parity.

 \begin{figure}
 \centering
 \includegraphics{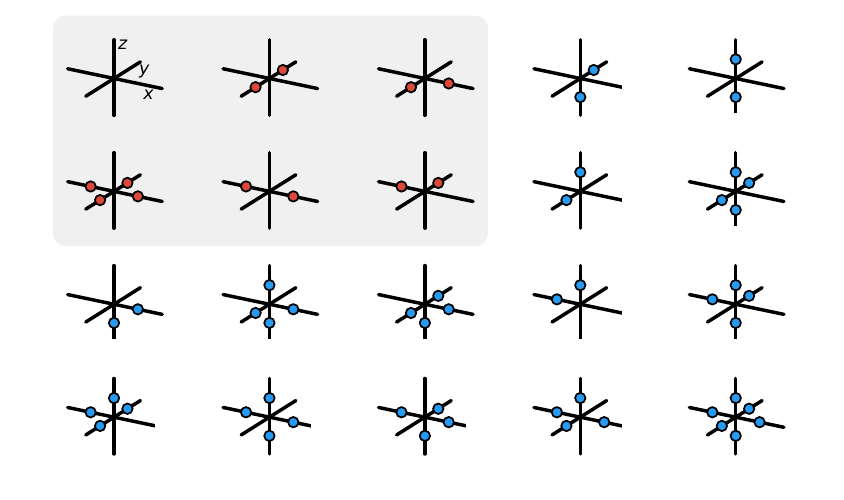}
 \caption{\label{fig:cartoon_3dgl} {\bf Gauss-law compatible states.} In total there are $20$
 allowed states for a 3D cubic lattice, and the grey-shaded area highlights the six compatible
 states for a 2D square lattice (where the $z$-component is neglected).}
 \end{figure}

 Using the infinitesimal generators $G_x$, one can generate a finite unitary transformation,
 $V = \prod_x \e^{-\i \theta_x G_x}$, where $\theta_x \in (0,2\pi]$ are local parameters. Then,
 under the gauge transformations, the spectrum and the eigenstates $\ket{\psi}$ remain unchanged,
 irrespective of their degeneracies:
 \begin{equation} \label{gauss2}
 \begin{aligned}
 H \ket{\psi} &= E  \ket{\psi} \Rightarrow V H V^\dagger \, V \ket{\psi} &= E \ket{\psi},
 \end{aligned}
 \end{equation}
 which follows from Eq.~(\ref{gauss1}). Note that any representation of the operators
 $E_{x,\mu}$, $U_{x,\mu}$, and $U^\dagger_{x,\mu}$ is admissible as long as the commutation
 relations in Eq.~(\ref{comm1}) are satisfied. The well-known case of Wilson-type lattice gauge
 theory uses the quantum rotor as a degree of freedom, generating an infinite-dimensional
 representation on each of the links. In this case, $U_{x,\mu}$ is an unitary operator, and 
 the commutation relation between $U_{x,\mu}$ and $U^\dagger_{x,\mu}$ vanishes. This is a
 special feature of the Wilson theory, which immediately narrows down the possible physical
 scenarios.

 Interestingly, a finite dimensional representation of the gauge fields can be obtained using 
 quantum spin-$S$ operators, $\vec{S}_{x,\mu}$. In particular, the raising and the lowering 
 spin operators can be identified with the quantum link gauge fields, and the z-component 
 with the electric field:
 \begin{equation} \label{spinrep}
   U_{x,\mu} = S^{+}_{x,\mu};~~U^\dagger_{x,\mu} = S^{-}_{x,\mu};~~E_{x,\mu} = S^z_{x,\mu}.
 \end{equation}
 Note that viewed this way, one way of approaching the Wilson limit of the gauge theory is to
 consider larger-spin representations~\cite{Schlittgen2001, Zache2021}.

 It is possible to give a pictorial representation of the QLM, which we show for the case of
 spin $S = \frac{1}{2}$. It is easiest to work in the electric flux basis, and the local
 Hilbert space is two-dimensional. We can represent left and right pointing arrows on the
 horizontal links as well as top and bottom pointing arrows on the vertical links to denote
 $\frac{1}{2}$ and $-\frac{1}{2}$, respectively. These considerations allow us to write the
 Hamiltonian of the $U(1)$ QLM in a more pictorial representation:
\begin{align}
  \label{eq:model_picture}
  \hat{H} &=  -J\sum_\square \left( \bigket{\flipA}\bigbra{\flipB} 
         + \bigket{\flipB}\bigbra{\flipA} \right) \nonumber \\
        &\qquad + \lambda\sum_\square \left( \bigket{\flipA}\bigbra{\flipA}
        +\bigket{\flipB}\bigbra{\flipB} \right).
\end{align}
 For the spin-$\frac{1}{2}$ case, the electric field energy terms contribute a constant and
 can be ignored. This corresponds to setting $g^2=0$, which we will consider for the rest of
 the article.

 It is instructive to point out that it is also possible to give a \emph{particle}
 interpretation of the spin-directions, such that $E = +\frac{1}{2}$ indicates the presence 
 of a hard-core boson, and $E = -\frac{1}{2}$ the absence of the particle. Then, the above
 pictorial Hamiltonian corresponds to
 \begin{align}
   \label{eq:model_picture_particle}
   \hat{H} &=  -J\sum_\square \left( \bigket{\flipAParticle}\bigbra{\flipBParticle} 
          + \bigket{\flipBParticle}\bigbra{\flipAParticle} \right) \nonumber \\
         &\qquad + \lambda\sum_\square \left( \bigket{\flipAParticle}\bigbra{\flipAParticle}
         + \bigket{\flipBParticle}\bigbra{\flipBParticle} \right).
 \end{align}
 This illustrates how Gauss' Law constrains the Hilbert space. For a hyper-cubic lattice, four
 links touch a site in two spatial dimensions, while six links touch a site in three spatial
 dimensions. Normally, this would have given rise to $2^4=16$ states in the former case, and
 $2^6=64$ states in the latter case. With Gauss' law, this would allow only six states in two
 dimensions, and 20 states in three dimensions. Their particle representation is sketched in 
 Fig.~\ref{fig:cartoon_3dgl}.

\subsection{Fermionic Quantum Link Model}
 Motivated by the particle representation, we introduce a new class of Abelian QLMs. This new
 class of models follows immediately from the particle formulation of QLMs in the previous
 section if one postulates that the particle is a fermion. This has the additional implication
 that the different link operators must also anti-commute, in addition to the first two
 commutation relations of Eq.~(\ref{comm1}), which are necessary for the gauge invariance of
 any microscopic model.

 Mathematically, we postulate that the two-dimensional Hilbert space at each link consists of
 two states: the absence or the presence of a fermion on the link. In the fermion occupation
 number basis, we can denote the two possibilities as 
 $\ket{0}$ and $\ket{1} = c^\dagger_{x,\mu} \ket{0}$, respectively. Here $c^\dagger_{x,\mu}$ 
 is a fermionic creation operator on the link joining the sites $x$ and $x + \hat{\mu}$.
 Similarly, $\ket{0} = c_{x,\mu} \ket{1}$, where $c_{x,\mu}$ is an annihilation operator. Since
 fermionic creation and annihilation operators anti-commute we have 
 $c_{x,\mu} c^\dagger_{x,\mu} = 1 - c^\dagger_{x,\mu} c_{x,\mu}$, and we can write
 $\ket{0} = c_{x,\mu} \ket{1} = c_{x,\mu} c^\dagger_{x,\mu} \ket{0} = (1 - c^\dagger_{x,\mu}
 c_{x,\mu}) \ket{0}$,
 so that we can interpret the number operator as $n_{x,\mu} = c^\dagger_{x,\mu} c_{x,\mu}$. 
 At this point, the similarities are obvious so that we  can identify the number operator as
 the electric field, and the creation and the annihilation operators as the quantum link and
 its Hermitian conjugate:
\begin{equation}
   U_{x,\mu} = c^\dagger_{x,\mu},\qquad U^\dagger_{x,\mu} = c_{x,\mu},\qquad E_{x,\mu} = n_{x,\mu}-\frac{1}{2}.
\end{equation}
 Note that with this identification the electric flux is still a bosonic operator, as is
 expected of a physical operator representing the electric field. The $1/2$ gives the electric
 flux the same values as a quantum spin $S = \frac{1}{2}$. The success of this novel
 identification of the operators is due to the fact that the creation and the annihilation
 operators satisfy the exact same commutation relations as the spin-$\frac{1}{2}$ operators:
 \begin{align} \label{fcommu}
  [n_{x,\mu}, c^\dagger_{y,\nu}] &= c^\dagger_{x,\mu} \delta_{x,y} \delta_{\mu,\nu},  \nonumber\\
  [n_{x,\mu}, c_{y,\nu}] &= c_{x,\mu} \delta_{x,y} \delta_{\mu,\nu}, \\
  [c^\dagger_{x,\mu}, c_{y,\nu}] &= 2 E_{x,\mu} \delta_{x,y} \delta_{\mu,\nu} = 
  2 (n_{x,\mu} - \frac{1}{2}) \delta_{x,y} \delta_{\mu,\nu}.  \nonumber
 \end{align}
 The quantum link operators themselves satisfy the anti-commutation relations:
 \begin{align}
 \{c_{x,\mu}, c_{y,\nu} \} &= \{c^\dagger_{x,\mu}, c^\dagger_{y,\nu} \} = 0,\nonumber \\
 \{ c^\dagger_{x,\mu}, c_{y,\nu}  \} &= \delta_{x,y} \delta_{\mu,\nu}.
 \end{align}
 The introduction of the fermionic operators is the key feature of this new class of QLMs. 
The fermionic worldlines have non-local correlations due to the Pauli exclusion principle, 
 and we expect qualitatively different phenomena to occur with fermionic links, beyond the
 ones realized in the bosonic version, and certainly beyond the ones in Wilson-type gauge
 theories.

 It is useful to note immediately that this proposed representation is very different from the
 rishon representations already motivated in \cite{Brower1999} and used in \cite{Banerjee2013a}
 for atomic quantum simulators. Note that the rishons are a generalization of the Schwinger
 boson construction, in which each link has a fixed number of fictitious particles called
 rishons, the number of which is determined by the representation. In an appropriately chosen
 basis, each quantum link operator essentially shifts the positions of the particles on a link.
 Additionally, there is an emergent link $U(1)$ gauge symmetry with the rishons, corresponding
 to the total number of rishons on a link. In contrast, the particle representation introduced
 here does not keep the total number of particles fixed within a link, but only globally.
 The particles, whether bosonic or fermionic, are free to move about on the lattice.

\begin{figure}[!tbh]
  \includegraphics[scale=1.0]{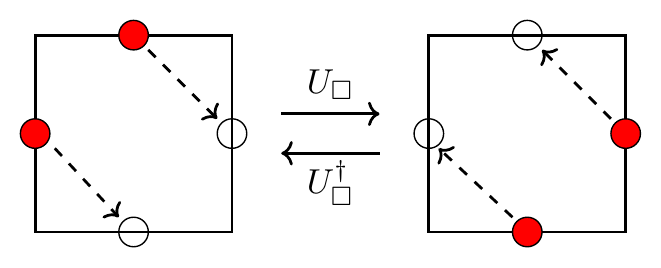}
  \caption{{\bf Correlated particle hop.} The plaquette involving fermionic operators can be
  understood as a simultaneous hop of the fermionic particles along the indicated lines.}
  \label{fig:plaq}
\end{figure}

In terms of the fermionic operators, we can now write the plaquette and local operators as:
 \begin{align}
 U_{\Box} &= c^\dagger_{x,\hat{i}} c^\dagger_{x+\hat{i},\hat{j}} c_{x+\hat{j},\hat{i}} c_{x,\hat{j}}, \nonumber \\
 G_x &= \sum_{i} \left( n_{x,\hat{i}} - n_{x-\hat{i},\hat{i}} \right).
 \label{eq:GLfermi}
\end{align}
We note that due to the anticommutation properties of the fermionic operators, the order of 
the operators matter and the theory will not be fully defined until the operator ordering for
the states in the Hilbert space is defined. The plaquette operator is composed of two creation
and two annihilation operators, and this works out to be a correlated hop of two fermions, as
shown in Fig.~\ref{fig:plaq}. This particular type of correlated hopping has peculiar
consequences, as will be explained in Sec. \ref{sect:Distinguish}. In particular, note that 
not all kinds of hoppings are possible, and this is the manifestation of the constraint,
consistent with Gauss' law. The only allowed hoppings are when shaded sites are occupied and
their directed neighbours are empty, in which case both the hoppings are oriented in the same
direction. Thus, among the six possible hoppings, only two are actually allowed. It can be
shown (via a unitary transformation) that the resulting theory is identical if the two 
hoppings instead occur in reverse directions.

\section{Methods} \label{sec:Methods}
In this section, we discuss the symmetries of the model under study as well as the employed
methodology. As mentioned previously, our main approach is the numerical diagonalization of
\equref{model_picture} on 3D lattices (with an even extent in all directions) up to $48$ 
links ($\gsystem{2}{2}{4}$). To this end, we employ the Lanczos algorithm~\cite{Lanczos1950} 
to extract a portion of the low-lying energy spectrum as well as the ground-state wavefunction.
Moreover, we discuss a systematic approximation, which we use for larger systems up to $96$
links ($\gsystem{2}{2}{6}$). Systems above $48$ links are out of reach for full diagonalization
for our current numerical implementation. A crucial ingredient for both approaches is to
efficiently construct the Hilbert space by finding all permissible Gauss' law states (GLS) on 
a given lattice. For completeness, we describe our algorithm in \appref{state_scaling}, where
we also briefly discuss the steep scaling of the number of GLS with lattice size.

\subsection{Symmetries of the microscopic model}
The Hamiltonian in \equref{model_picture} features several global symmetries, which we may use
to reduce the computational burden by separately diagonalizing the blocks corresponding to the
different quantum numbers associated with the symmetries. Below, we explain these symmetries
and how they are implemented for the bosonic and the fermionic representations.

The overwhelming advantage arises from exploiting the conservation of the winding numbers in a
given plane under plaquette flips in this plane. On a 3-d lattice, there are three such
separately conserved winding numbers, generating a $U(1)\otimes U(1) \otimes U(1)$ symmetry 
and mathematically expressed as:
\begin{align}
  W_x &= \frac{1}{L_{y} L_{z}} \sum_{i, x=x_0} E_{i, \hat{x}}, \\
  W_y &= \frac{1}{L_{x} L_{z}} \sum_{i, y=y_0} E_{i, \hat{y}}, \\
  W_z &= \frac{1}{L_{x} L_{y}} \sum_{i, z=z_0} E_{i,\hat{z}}.
\end{align}
For a pictorial representation of the winding numbers on a $\gsystem{2}{2}{2}$ lattice see 
\figref{link_numbering}. In the particle representation, recall that the $E_{i,\mu}$ operators 
should be replaced  by $n_{i,\mu} -\tfrac{1}{2}$, since the occupation numbers can be only 
0 or 1. This puts the E-flux values to be in one-to-one correspondence with the original 
formulation using spin-$\frac{1}{2}$. In this notation, the winding number in each direction 
can go from $-L_\mu/2$ to $L_\mu/2$, thus, resulting in a total of $(L_{\rm x} + 1) \times 
(L_{\rm y} + 1) \times (L_{\rm z} + 1)$ sectors for a lattice with even 
$(L_{\rm x}, L_{\rm y}, L_{\rm z})$ extents. For any lattice, the $W = [W_x~W_y~W_z] = [000]$ 
winding sector will be the largest block and is a-priori expected to host the ground-state, 
being the most symmetric configuration.

  The Hamiltonian has a $\mathbb{Z}_2$ charge-conjugation symmetry. The unitary transformation
 is implemented on an operator ${\cal O}$ as $^C {\cal O} = C {\cal O} C^\dagger$. In the
 original formulation with quantum spins, this yields $^C U = U^\dagger;~~^C U^\dagger = U;
 ~~^C E = -E$. For the spin representation (which is equivalent to the hardcore boson
 representation), the charge conjugation operator is $C = \sigma^x$. In the fermion
 representation, in terms of the creation and annihilation operators acting on individual links
 we have, $C = (\cre{} + \i \ani{})$. Using the fermion anti-commutators (or the Pauli-matrices
 for the spins), it is easy to show that $C^\dagger C = 1 = C^2$. For the fermions, additional
 phases are involved in the transformation of the individual link operators, but the observable
 electric flux transforms as in the spin-representation:
\begin{align}
   & ^C \cre{} = \i \ani{};\qquad ^C \ani{} = -\i \cre{}; \nonumber \\
   &  ^C (n-\frac{1}{2})  = -(n - \frac{1}{2}).
\end{align}
To define the transformation on the entire system, or on the wavefunction, an ordering of the
links on the lattice needs to chosen and the product of individual transformations taken along
the ordering:
\begin{equation}
  C = \prod_{i=1}^{N_s}\left[\cre{i} + \i \ani{i} \right].
\end{equation}
For the spin representation, this works out to be a product of $\sigma^x$ on all links, while
for the hardcore bosons use bosonic creation and annihilation operators which commute for
unequal sites, but anti-commute for identical sites \cite{Matsubara1956}. Obviously, the ordering 
is not important for the spins or the hardcore bosons. At the single-plaquette level, the
hermitian conjugate of the plaquette-flip operator $U_\square$, is $U_\square^\dagger$, and is
identical to the charge conjugation operation. Similarly, it is easy to see that in the absence
of any matter, Gauss' law is also satisfied under charge conjugation. Note that this
transformation preserves the commutation relation of the quantum spins, as well as the
anti-commutation relations of the fermions.

Similarly, the Hamiltonian is invariant under parity transformations, which can be defined as
for the fermionic representation:
\begin{equation}
  P = \prod_{i=1}^{N_s}\left[\cre{P(i)} + \i \ani{P(i)} \right]
\end{equation}
where $P(i)$ simply denotes the point reflected index around the origin (with appropriately
imposed periodic boundary conditions). Note that under the parity operation, the links and 
the flux transforms as: 
$^P U_{xy} \rightarrow U^\dagger_{-y,-x};~ ^P E_{xy} \rightarrow -E_{xy}$.

 In addition, the model has the other point group symmetries, such as translation invariance
 (in each of the x-, y-, and the z-directions), the rotation symmetries (the $C_4$ rotations
 about the lattice axes, the $C_3$ rotations about the body diagonals, the $C_2$ rotations
 about the axis joining the opposite edges; the subscript $n$ denotes the $2 \pi/n$-fold
 rotation. While it is possible to take advantage of the commuting symmetries to increase the
 numerical reach of our exact diagonalization routines, we have not considered it here.

\subsection{Low-energy approximation}
 Because of the prohibitive scaling of the number of GLS with system size (see
 \appref{state_scaling} for a discussion), ED is restricted to the lowest system sizes (for us
 these are  $\gsystem{2}{2}{2}$ and $\gsystem{2}{2}{4}$, which involve 24 and 48 links
 respectively). For larger volumes, the number of states requires serious numerical effort at
 the limit (or beyond) what is currently feasible on HPC setups. Moreover, in higher
 dimensions, increasing the linear dimension by a unit amounts to increasing the total number
 of links proportional to the surface area.

  To gain some insight into the physics despite these limitations, we construct a truncated
  Hilbert space (sometimes also called a limited functional space) starting from the
  energetically lowest lying states and systematically introducing excitations to form new
  basis states (similar strategies have been applied in ED-like studies for various other
  physical systems, see e.g.~\cite{Bonca2007,Grining2015}). In the present case, the
  energetically most favourable states at large negative $\lambda$ are the ones with the most
  flippable plaquettes. An excitation can then be introduced by flipping single plaquettes,
  which constructs a new state while respecting Gauss' law. Exhausting all maximally-flippable
  states in this way, one obtains a set of states that differ by a single flip from the
  lowest-lying states, in the following denoted as ``flip-level'' $1$ (FL1). Higher FLs are
  reached by repeatedly applying this procedure to the newly found states. In fact, this is an
  alternative method to construct the full list of GLS. However, a tiny subset of non-flippable
  states are omitted in this way and this procedure may be stuck in disjoint pockets of the
  Hilbert space, which is sometimes called fragmentation of Hilbert space in the literature
  \cite{Sala2020,Moudgalya2021,Mukherjee2021}. Nevertheless, as we will see, the unflippable
  states are not of interest for the current study, and further no Hilbert-space fragmentation
  is present for the square lattice Hamiltonian.

  While it is obvious that eventually the spectrum of the truncated Hilbert space will converge
  to the true spectrum, we expect that the convergence sets in early such that we may extract
  useful information about the finite-size scaling of the mass gap. As it turns out, this
  approach is feasible and allows us to study the physics of the lattice $\gsystem{2}{2}{6}$,
  having $72$ links, without the need to fully diagonalize the entire Hamiltonian. Further
  details, including the convergence analysis for the $\gsystem{2}{2}{4}$ system, are shown in
  \appref{low_energy_details}.

\section{Distinguishing the Bosonic and Fermionic Quantum Link Models} \label{sect:Distinguish}
  The $U(1)$ quantum link model in $(2+1)-$d has been extensively studied on the square lattice
  in the spin-$\frac{1}{2}$ representation \cite{Banerjee2013,Tschirsich2019}. Therefore, we
  begin our investigation with the $(2+1)-$d fermionic model, attempting to understand if it
  has different properties from the one realized with quantum spins. We further examine the
  spin and fermionic versions of the model in $(3+1)-$d for similarities and differences.

\subsection{Two Dimensions}
\begin{figure}
  \includegraphics{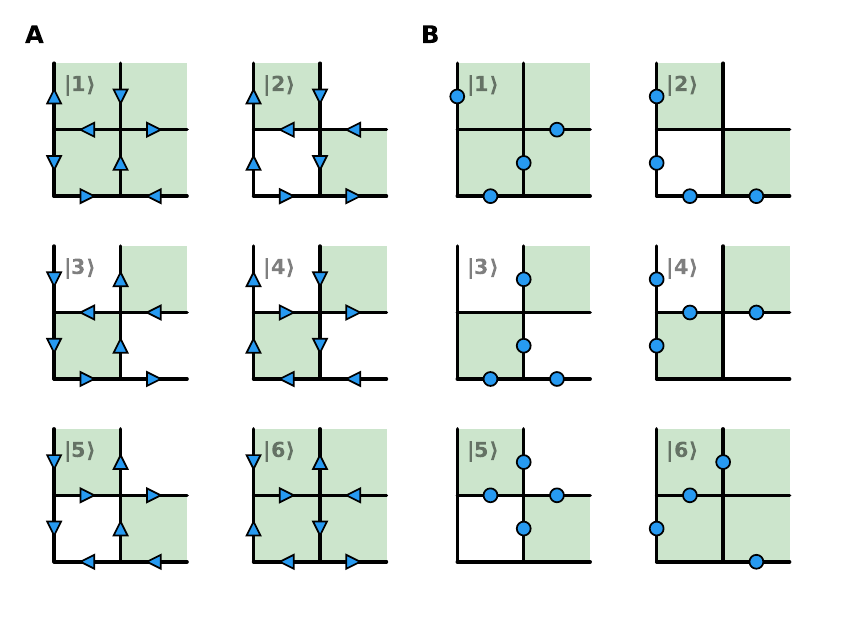}
  \caption{\label{fig:cartoons_2d} {\bf Cartoon states of the zero-winding sector} for a 
  $2 \times 2$ lattice in the  spin (A) and particle (B) representation. Green (red) shaded
  plaquettes are flippable (non-flippable). }
\end{figure}

\begin{figure}
  \includegraphics[width=\columnwidth]{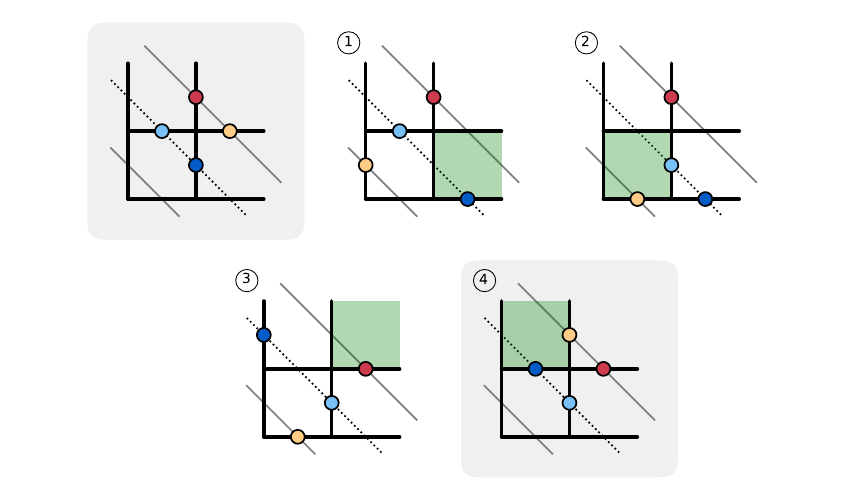}
  \caption{\label{fig:fermi_2d} {\bf Restricted movement on the 2D lattice.} The nature of the
  plaquette interaction constrains the ``paths'' of the fermionic particles in two dimensions
  along diagonal tracks, illustrated by the lines in this particular $2 \times 2$ example.
  Effectively, the plaquette interaction causes two correlated simultaneous hops along two
  adjacent 1-dimensional chains. In the numbered figures, the green highlighted plaquettes
  illustrate the particular Hamiltonian term being applied in each step. The only sign for the
  fermionic case that could occur is due to the boundary, but the fact that the particles move
  in pairs ends up precluding this possibility.}
\end{figure}

  In order to understand the differences and the similarities between the fermionic and bosonic
  representations of the lattice gauge theory, we begin by considering the simplest possible
  setting: the case of the $2 \times 2$ lattice with periodic boundary conditions. For this
  system, there are 4 sites, and 8 links. Implemented without any further constraints, the
  system would have $2^8 = 256$ states, but imposing Gauss' law of $Q_x = G_x = 0$ for every
  site gives rise to only 18 total states of the system. The corresponding states (in the zero
  winding sector) in both the spin and the fermion representations are shown in panels (A) and
  (B) of \figref{cartoons_2d}, respectively.

  The action of the Hamiltonian for both the bosonic and fermionic models (when $\lambda=0$) 
  on the different states, as numbered in \figref{cartoons_2d}, is then given by:
\begin{equation}
  \begin{aligned}
    H \ket{1}  &= -J( \ket{2} \pm \ket{3} + \ket{4} \pm \ket{5}) \\
    H \ket{2}  &= -J(\ket{1} \pm \ket{6} ) \\
    H \ket{3}  &= -J( \pm \ket{1} + \ket{6}) \\
    H \ket{4}  &= -J( \ket{1} \pm  \ket{6}) \\
    H \ket{5}  &= -J(\pm \ket{1} + \ket{6}) \\
    H \ket{6}  &= -J(\pm \ket{2} + \ket{3} \pm \ket{4} + \ket{5}).
    \label{actions}
  \end{aligned}
\end{equation}

 (Details are given in App.~\ref{app:diag}). In the bosonic case the upper (positive) signs 
 in Eq.~(\ref{actions}) are taken, and in the fermionic case the lower (negative) signs are
 taken. The spectrum obtained in the two cases, however, is identical.

 Naively this seems surprising, since from the analysis of spin and fermionic models one knows
 that in the latter, the fermionic world lines can exchange positions in two spatial
 dimensions, which gives rise to different physics as compared to the bosonic version.
 Therefore, it must be that the nature of the four body interactions, necessary to preserve
 gauge invariance also forbids all those paths which could otherwise differentiate between the
 hard-core bosons and the fermions. A geometric proof of this is provided in \figref{fermi_2d},
 which can be easily extended to any square (or rectangular lattice) with linear dimension $L$.
 Note further, that the proof can be extended for all the superselection sectors labelled by
 different values of charges $Q_x$. We have explicitly repeated the exercise on the $2\times 2$
 lattice for all possible values of the fermion occupation (i.e., without imposing the Gauss
 Law), and obtained an identical spectrum for both the spin-links and the fermion-links. This
 implies that the physics of the fermionic model is also the same as the ones already studied
 before.

\subsection{Three Dimensions}
\begin{figure*}
  \includegraphics[width=0.9\textwidth]{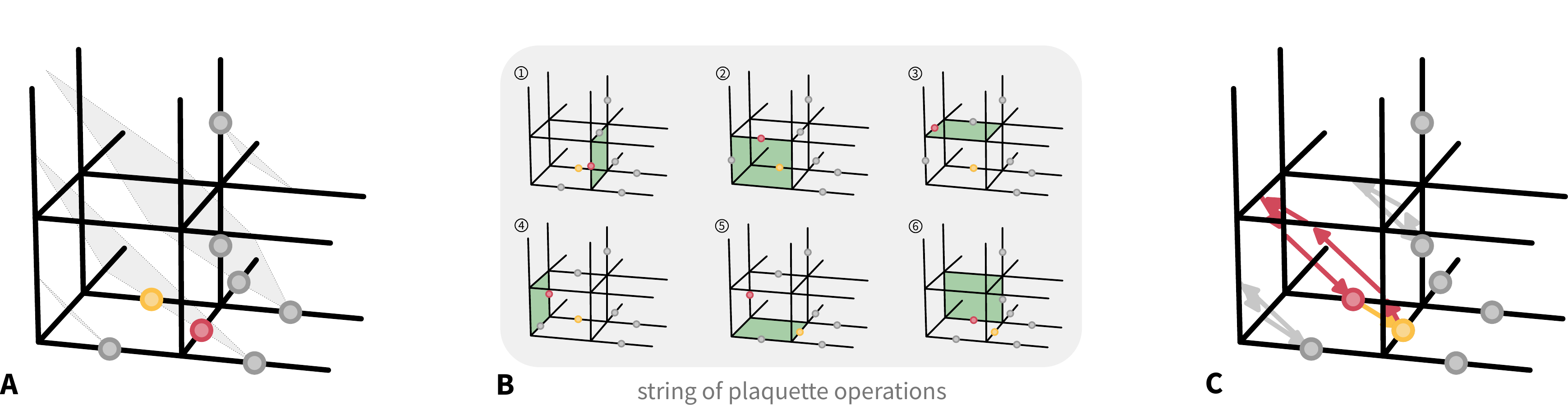}
  \caption{\label{fig:fermi_exchange} {\bf Difference between bosonic and fermionic links.}
  Starting from a given initial position on a $\gsystem{2}{2}{2}$ lattice (A) one may follow a
  ``path'' of two fermions by successively applying suitable plaquette flipping operators (B,
  application from top left to bottom right). Finally, one ends up in the initial configuration
  with an effective interchange of two fermions (C), implying a sign that is not present in the
  bosonic model.}
\end{figure*}

 In three spatial dimensions, we can extend the geometric proof outlined for two spatial
 dimensions. In this case, the particles are restricted to move on planes, and it is possible
 for a particle to make an orbit without disturbing any other particles in the plane. To see
 this, consider the example depicted in \figref{fermi_exchange}: Starting from an initial
 configuration (A), one may apply a string of plaquette operators that move the particles
 according to the paths shown in panel (B). In the final configuration (C), all links returned
 to their original occupation number with an effective interchange of the colored particles
 (the exact algebra for the applied operators is carried out in \appref{fermi_exchange}). 
 While in the bosonic case there is no sign associated with this exchange, the fermionic nature
 of the gauge particles requires a sign, therefore the physics of both models are expected to
 exhibit distinct effects.

 Thus, we obtain the very interesting result that interactions responsible for maintaining
 gauge invariance restrict worldlines which cause particles two swap positions on the square
 lattice, thus rendering the statistics of the particles irrelevant. However, in three spatial
 dimensions, this is no longer the case, the particles move along planes, and the bosonic and
 the fermionic physics differ, since the worldlines which give different signs to the bosons
 and fermions can occur. Interestingly, the subdimensional motion of particles observed in this
 model due to the gauge interactions is reminiscent of fractonic physics
 \cite{Nandkishore2019}.

\section{Physics in three dimensions}
 In this section we numerically explore the physics of both the spin-$1/2$ bosonic and
 fermionic versions of the QLM in 3D, which we have shown to have distinct worldline weights
 to each other, in contrast to in 2D. Another difference compared to the 2D system is that in
 3D there is no configuration on the lattice that allows all plaquettes to be flippable
 simultaneously, regardless of the statistics of the gauge particles. As we shall see below,
 this leads to a different broken symmetry in the ordered phase expected at a large negative RK
 coupling $\lambda$ on 3D lattices as compared to 2D. We present results for several
 observables to explore both the bosonic and fermionic models, emphasizing the different
 features found in both cases.

\subsection{Spectrum vs. $\lambda$}  \label{sect:spectrum}
\begin{figure}[h!]
  \includegraphics[width=\columnwidth]{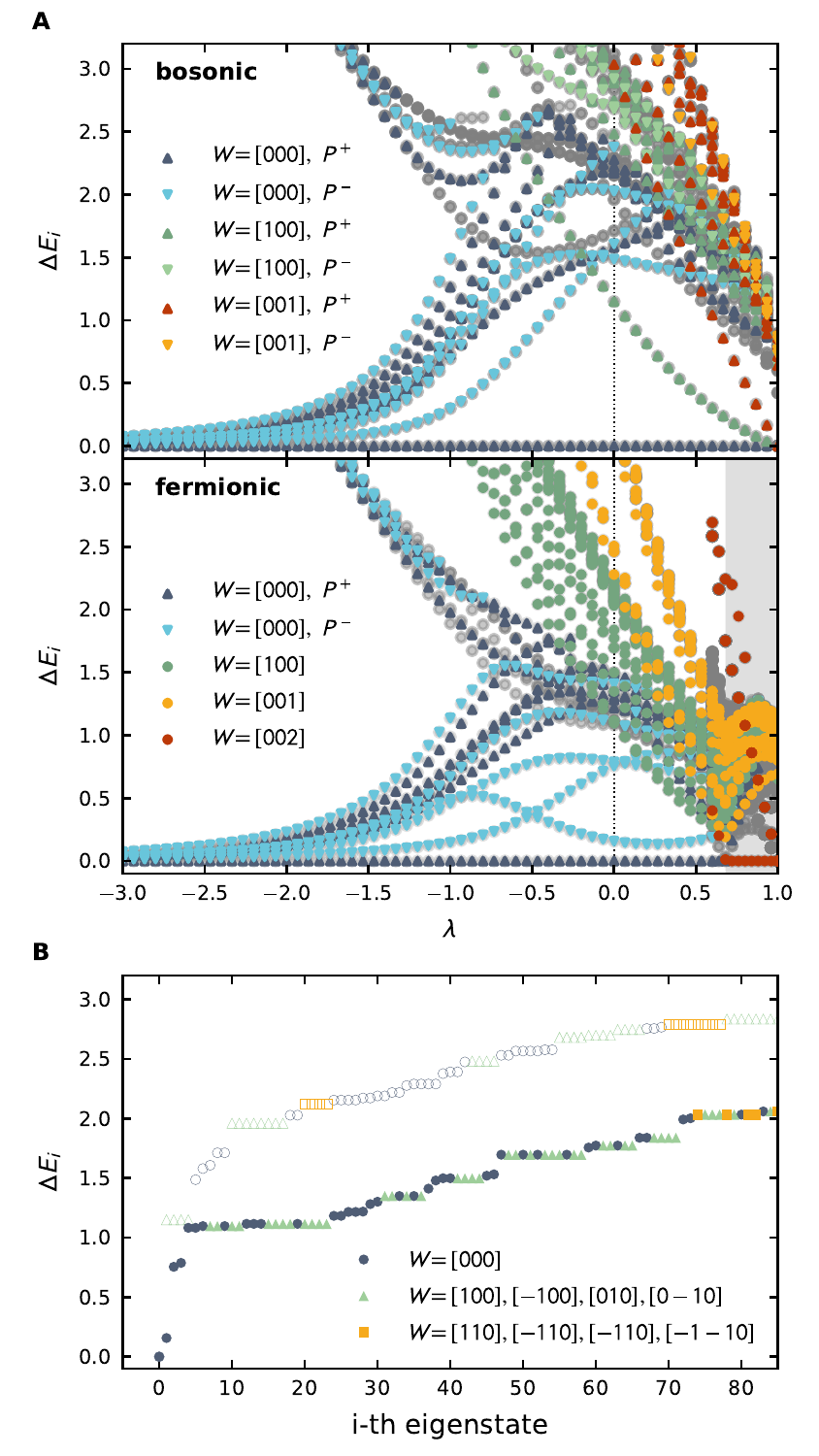}
  \caption{{\bf Low energy spectrum}  for $\gsystem{2}{2}{4}$ lattices with bosonic (A, top
  panel) and fermionic (A, bottom panel) links. For fermions, the gray shaded area marks the
  region where the ground-state is in the non-zero winding sector. (B) Lower part of the 
  spectrum for bosons (open symbols) and fermions (solid symbols) at $\lambda=0$, 
  corresponding to the dotted vertical lines in panel A.}
  \label{fig:spectra}
\end{figure}
 As our first quantity of interest, we study the low-energy spectrum and its dependence on 
 the RK coupling $\lambda$. In panel A of \figref{spectra}, the spectra for both bosonic (top)
 and fermionic (bottom) links are shown. First, we note the similarity of the low-energy
 spectra of both models at large negative $\lambda$, which persists up to $\lambda \lesssim
 -1.0$. This comes as no surprise, since in the limit of $\lambda\to-\infty$ the particle
 statistics, i.e. the plaquette flipping term in the Hamiltonian, does not play any role and
 arguments may be made purely based on energetic considerations. Hence, in this limit the 3D
 system wants to maximize the number of flippable plaquettes. However, as mentioned above, not
 all plaquettes can be made flippable at the same time--only $2/3$ of the total number
 of plaquettes $N_p = 3\times L_x \times L_y\times L_z$. The maximally flippable configurations
 are achieved by stacking fully flippable layers in a given planar direction while maximizing the number
 of flippable plaquettes along the remaining two planar directions. For a given direction, this stacking
 while retaining the maximal number of flippable plaquettes can be done in four ways, such that
 in total there are $4 \times 3 $ most flippable configurations which correspond to the 
 ``half-filled'' or particle-hole symmetric case (in the spin picture this would be $S_z = 0$).
 Therefore, at $\lambda \to -\infty$ we observe a $12$-fold degeneracy in the spectrum. It
 turns out that these states are related via the transformation of the group $D_{3h} = D_3
 \otimes \mathbb{Z}_2$, which represents the direct product of the lattice rotation in 3D 
 and the charge conjugation, respectively. All states belonging to this multiplet are expected
 to be degenerate in the thermodynamic limit (TL), and therefore this symmetry is spontaneously
 broken in this phase.

\begin{figure*}
  \centering
  \includegraphics[width=\textwidth]{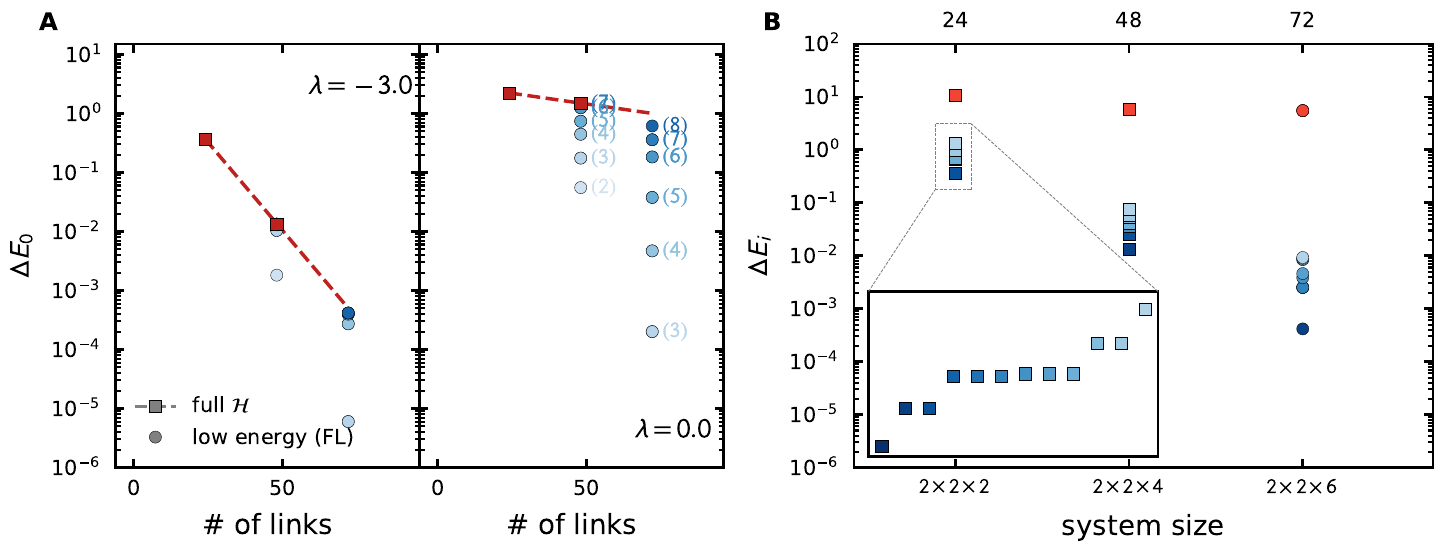}
  \caption{\label{fig:gap12} {\bf Finite volume behavior of the low-energy spectrum for bosonic links.}
  (A) Finite size scaling of the gap to the first excited state on a $y$-log scale. Fast decay for 
  $\lambda=-3.0$  (left panel) versus  a slower decay for $\lambda = 0$ (right panel). (B) Gaps to 
  the lowest $11$ excited states are shown in blue shades, higher excited states in red. The scale 
  on the top denotes the corresponding number of links. (Inset) Degeneracy structure of the ground-state 
  manifold on a $\gsystem{2}{2}{2}$ lattice. In all panels, squares denote results from a diagonalization 
  of the full Hilbert space $\mathcal{H}$, whereas circles show low-energy results from a reduced Hilbert 
  space with the level of approximation indicated in parentheses (see main text). Dashed lines are a 
  guide to the eye. }
\end{figure*}

  This observation is further elucidated in panel B of \figref{gap12}, where the volume scaling of the
 lowest few energy gaps in the bosonic model is shown for a reference value of $\lambda=-3.0$,
 which is deep in the symmetry-broken phase. The spectrum for the fermionic model is virtually 
 indistinguishable from the bosonic model in that regime, and so the low-energy spectra
 for both models differ by less than one percent. As expected from the above arguments, the 
 lowest $11$ gaps decay
 exponentially while higher-lying energy values decay more slowly, seemingly polynomially.
 Moreover, the inset of the same figure shows the lowest $12$ states for a $\gsystem{2}{2}{2}$
 lattice, revealing a structure with six energy manifolds consisting of $1+2+3+3+2+1$ states.
 This corresponds exactly to the dimensionality of the irreducible representations of
 the $D_{3h}$ group, where the (numerically) exact degeneracies are a consequence of the
 non-Abelian nature of this group.

At larger values of $\lambda$, the competition between maximizing flippability and the kinetic 
term in \equref{model_picture} may potentially lead to a phase transition out of the ordered phase.
Earlier studies of dimer models (with bosonic links) on other 3D lattice geometries
~\cite{Sikora2009,Sikora2011} suggested the emergence of a quantum spin liquid (QSL) phase that 
persists at intermediate $\lambda$ values up to the RK point at $\lambda = 1.0$. One of the hallmarks 
of such QSLs is the absence of any sort of symmetry breaking in the ground state in the thermodynamic
limit, which in itself is challenging to establish rigorously. A sense of this can already be 
obtained from panel B of \figref{spectra}, which shows the low-energy spectra of both the 
bosonic and the fermionic QLMs in different winding sectors. We note the presence of a large
number of low-lying energy eigenstates above the ground state for both the cases, without the 
presence of any large gap. This indicates the possibility of smooth excitations which have overlaps 
with the low-lying eigenstates, and consequently a smooth spectral function lacking
any distinctive energy scale, which is typical of a liquid phase. This is in sharp contrast to the nematically 
ordered phase, where above the manifold of 12 states there is a large window which 
hosts no energy eigenstates. Consequently, excitations there are peaked around a certain
frequency and typical of a symmetry broken solid phase. We have noted that this distinction
between the two regimes persist for our system sizes, and therefore if the trend continues to 
the TL, a liquid phase with continuous excitations is reasonable in the $\lambda \sim 0$ region.

 Moreover, one may investigate the volume scaling of the gap between the ground and the first 
excited state, where the absence of any exponential decay of the gap, $\Delta E_0$, would be 
indicative of such an liquid phase around $\lambda \sim 0$. In panel A of \figref{gap12}, 
we show results for the lowest energy gap for bosonic link variables for two representative 
values of $\lambda$. The left panel at $\lambda = -3.0$ again reflects the symmetry broken 
phase (c.f. panel B of the same figure) where the gap to the first 
excited state decays exponentially with the volume, as one would expect for a symmetry broken phase. 
Conversely, the right-hand panel of the same figure corresponds to $\lambda = 0$ and displays a much 
slower decay. Moreover, we note that in the $\lambda=0$ phase, the lowest excitation is a winding string 
in the shorter directions $[100]$ and $[010]$, while the lowest excitation within the zero-winding sector 
costs more energy. This is an interesting indication that the vacuum is stable to the creation of strings,
another signal for a possible Coulomb phase. The values of $\lambda$ were chosen solely to be
representative of the respective regimes, and their numerical values do not correspond to 
quantitatively important aspects of the phase transition. The data points for the largest system, 
here $\gsystem{2}{2}{6}$, have been obtained with the low-energy approximation discussed above 
and are in agreement with this trend.

 While this analysis presents some evidence for a transition between an ordered phase and a 
potential QSL, the obtainable lattice sizes limit our ability to make definitive statements. 
Moreover, the exact transition point (if present at all) is impossible to pinpoint with ED 
with the few lattice sizes that we possess. A natural next step would be to address these 
issues by means of ab-initio calculations, such as a Markov chain Monte Carlo. However, for 
the fermionic model, sign problems currently preclude the existence of any efficient Monte-Carlo
algorithm to our knowledge, and different numerical approaches such as tensor network methods
\cite{Felser2021,Tepaske2021,Magnifico2021} might be necessary.

  While the volume scaling of the gap is here only presented for bosonic link variables, we 
observe qualitatively similar behavior for fermionic links and therefore expect the transition 
between an ordered phase and a QSL for both models (we return to this point in \sectref{histograms}).
However, despite the similarities of the low-energy spectrum in both models, there are two 
important differences. The first observation concerns the winding numbers of the first excited 
state in the potential QSL phase, which seem to be nonzero only for the bosonic case. The model 
with fermionic degrees of freedom has another state of the zero-winding sector as first excited
state (although with different parity as the GS). Only upon increasing $\lambda$ further do we 
observe the eigenstates with non-zero winding approaching the ground state.

 Interestingly, there is a second key difference between the models in that the fermionic model 
even features a ground-state level crossing at $\lambda_{c_2} \approx 0.65$ for the lattice 
$2 \times 2 \times 4$, which is absent for the bosonic links. Beyond this $\lambda_{c_2}$, and 
up to the RK point, the GS is in the $W_x=W_y=0$ and $W_z=2$ winding sector, indicated by the
gray-shaded area in panel A of \figref{spectra}. Therefore, the system seems to enter a ``flux-condensed'' 
phase where it is energetically advantageous for the system to have flux-lines along the long 
direction of the tube-like lattice. It will be interesting to see whether this feature persists 
in the TL.

\subsection{Ground-state fidelity susceptibility}
 In order to further shed light on whether the explored parameter range of $\lambda$ crosses a phase
transition, one can also exploit tools from quantum information theory. Specifically, in this section
we present results for the ground-state fidelity, which measures the overlap between two ground states
with a slight difference in the coupling. The overlap exhibits a dip if the two corresponding ground
state (GS) wavefunctions are qualitatively different, i.e., when the states belong to different quantum
phases, and hence is a useful witness to detect quantum phase transitions~\cite{Zanardi2006}. Here, 
we are interested in the GS fidelity as a function of $\lambda$, defined as
\begin{equation}
  F(\lambda,\epsilon) = |\langle\psi(\lambda)|\psi(\lambda+\epsilon)\rangle|.
  \label{eq:fidelity}
\end{equation}
where $\epsilon$ denotes the difference between the two involved $\lambda$ values. Note that the
fidelity itself depends on $\epsilon$ since, naively, the overlap between two ``neighbouring'' 
states increases when the parameter offset is reduced (since the states are ``closer'').
Moreover, this overlap is expected to vanish exponentially with increasing system size, in 
accordance to predictions of random matrix theory \cite{Alessio2016}. To overcome these 
shortcomings, it is more convenient to introduce the fidelity susceptibility formally defined via
\begin{equation}
  \chi_{\rm F} \equiv -\frac{\partial^2\log{F}}{\partial\epsilon^2}\bigg|_{\epsilon=0}.
\end{equation}
 Much like thermodynamic susceptibilities~\cite{You2007}, the fidelity susceptibility encodes the
response of the GS overlap to a small change in the driving parameter $\lambda$. Moreover, just 
like the fidelity of order parameters, the positions of the maxima in $\chi_{\rm F}$
at finite system sizes can be used to extract critical properties for infinite systems via 
finite-size scaling~\cite{You2007,Wang2015}.

 Computationally, there are several ways to extract the fidelity susceptibility (see, e.g.,
Ref.~\cite{Wang2015} for an overview). A straightforward way is to simply take the overlap 
according to \equref{fidelity} and then exploit the relation
\begin{equation}
  F(\lambda,\epsilon) = 1 + \frac{\epsilon^2}{2}\chi_{\rm F}(\lambda) + \mathcal{O}(\epsilon^4).
\end{equation}
which may be obtained via first-order perturbation theory~\cite{You2007,Gu2010}. A potential 
drawback is a systematic error induced by the finite difference $\epsilon$, however, this can be
efficiently suppressed by using small enough $\epsilon$.
\begin{figure}[b]
  \includegraphics{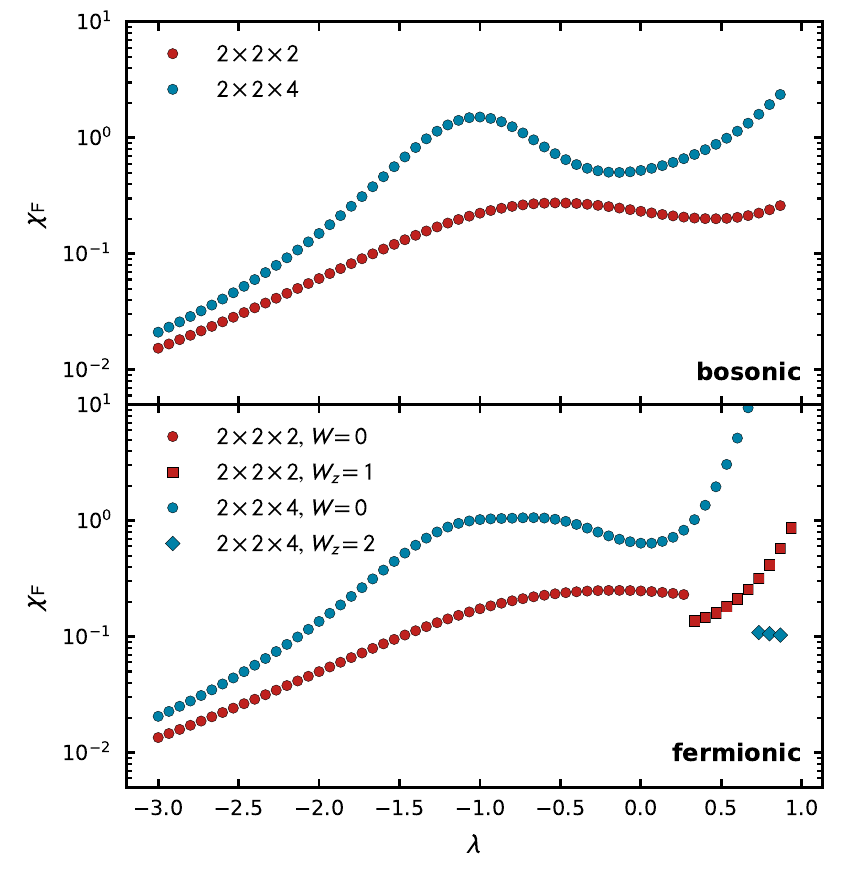}
  \caption{{\bf Ground-state fidelity susceptibility} for two system sizes for bosonic (top) and
  fermionic (bottom) links.  Note the log-scale on the $y$-axis.}
  \label{fig:gs_fidelity}
\end{figure}

 In \figref{gs_fidelity} we show results for the fidelity susceptibility on both lattice sizes 
studied in this work for bosons (top) and fermions (bottom). For the bosonic link model we observe 
a peak for $\lambda \sim -1$, hinting at the presence of a smooth phase transition, and thereby
suggesting that for $\lambda$ close to zero, the system probably comes out of the ordered phase. 
At positive $\lambda$ the susceptibility dips before a steep increase heralds the presence of 
the RK point at $\lambda = 1$ (which separates a potential QSL from a staggered dimerized ground 
state with no flippable plaquettes). Close to the RK point, the ground state is sensitive to 
states with winding strings, and this shows up as a sharp increase in $\chi_{\rm F}$.

 For the case of fermionic links, the overall behavior is similar, however, a qualitative 
difference arises: while the sharp rise in the vicinity of the RK point as well as the smooth 
peak at $\lambda \sim -1$ are present as in the bosonic case, the curves for both shown 
system sizes exhibit a discontinuity. Unsurprisingly, this jump occurs at the point where
we observe the ground-state level crossing in the spectrum, i.e., where a potential 
first-order transition takes place. For both lines shown in the lower panel of 
\figref{gs_fidelity}, the solid circles are the values in the zero-winding sector whereas 
the filled squares reflect the susceptibility in a non-zero winding sector, which hosts the 
ground-state in this regime.

 Based on these findings and the discussion in the previous section, we propose the qualitative
phase-diagram on the $\lambda$-axis shown in \figref{pd_sketch}. For both the bosonic and 
fermionic link models, there seems to be a smooth quantum phase transition at moderately 
negative $\lambda$. Additionally, for fermionic links a sharp transition to an as-of-yet 
unknown phase seems to exist, however, it is challenging to predict if this persists in the
thermodynamic limit. In the next section, we investigate other observables to establish 
some properties of the putative quantum phases.

\subsection{State participation and entropy\label{sect:histograms}}
\begin{figure*}[t]
  \centering
  \includegraphics[width=\textwidth]{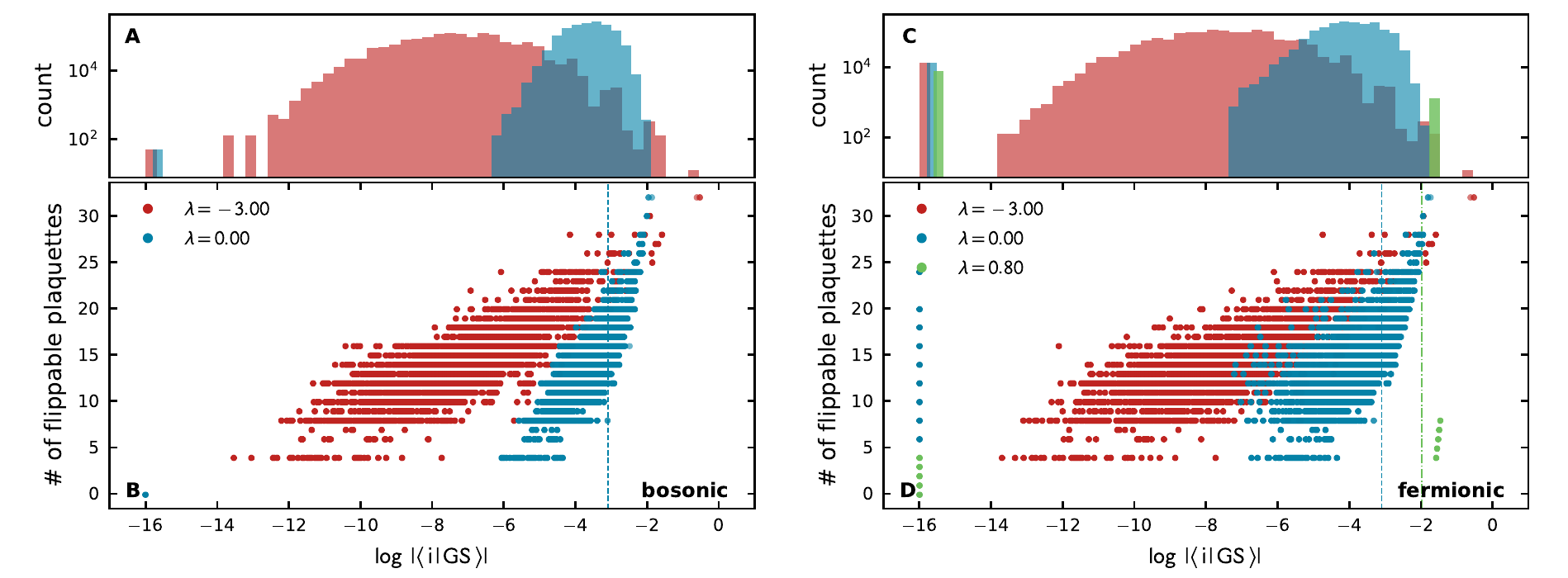}
  \caption{\label{fig:state_histograms} {\bf Correlations between the weight and the number 
  of flippable plaquettes} of basis-states in the ground-state for bosonic (left) and 
  fermionic (right) links on a $\gsystem{2}{2}{4}$ lattice. (Top row) Histogram of the
  log-values of the absolute value of the weight. At large negative $\lambda = -3.0$ only 
  a few states contribute significantly, others have significantly reduced weight and are 
  spread over orders of magnitude. At $\lambda =0$ the distribution is narrow, corresponding 
  to a more even distribution of weight among the states. (Bottom row) The number of flippable
  plaquettes vs. the log of the absolute value of the amplitude. The dashed line marks an even
  superposition of all states.}
\end{figure*}

 In order to further characterize the ground-state behavior, we investigate the structure of 
 the ground-state wavefunction at representative values of the RK coupling $\lambda$ in the 
 different phases. As a first step, we discuss a correlation histogram that relates the number of 
 flippable plaquettes for a given basis state with the relative weight of this basis state in 
 the GS wavefunction on a $\gsystem{2}{2}{4}$ lattice, shown in \figref{state_histograms} for 
 both the bosonic and the fermionic links.

 While both scenarios look fairly similar, let us first discuss the case of bosonic links (left column). 
 At a representative value for the rotational symmetry broken ordered phase, for which we take 
 $\lambda = -3.0$, the ground-state wavefunction $\gs{-3.0}$ is in the $W = [000]$ sector (red symbols). 
 Both panel A and panel B show the absolute value of the overlap of the i-th basis state with the 
 ground state wavefunction: $c_i = \overlap{i}{\mathrm{GS}}$. The y-axis of panel A shows the number 
 of basis states which have the same overlap with the wavefunction, while panel B shows the total 
 number of flippable plaquettes for the corresponding basis states. As is apparent from panel B, 
 almost the entire weight in this regime is carried by basis states with the largest number of 
 flippable plaquettes. This is indicated by the isolated scatter points in the top right corner. 
 Indeed, this is not unexpected, since we already argued above that in this phase the $D_{3h}$ 
 symmetric states constitute the GS manifold at $\lambda\to-\infty$. The relative importance of 
 these few states becomes even more apparent by considering the corresponding histogram of the 
 weights of the GS wavefunction (not resolved in the number of flippable plaquettes), shown in 
 panel A of \figref{state_histograms}. Here, the isolated point corresponds to the single peak 
 with the largest weight and all the other basis states contribute with weights smaller by several 
 orders of magnitude (note the logarithmic scale).

  At larger values of $\lambda$, we have argued based on the scaling of the energy gap to the first 
 excited state, that symmetry breaking might be absent, and the system enters a disordered QSL phase. 
 Performing an equivalent analysis on the representative wavefunction at $\lambda=0.0$ (blue data) 
 indeed reveals a picture consistent with the conjecture of a liquid phase: The amplitude in $\gs{0.0}$ 
 is carried by many states with vastly different number of flippable plaquettes, all with similar 
 amplitudes. This is reflected by the relatively flat distribution in panel B of 
 \figref{state_histograms} and by the narrow histogram in panel A of the same figure. Moreover, 
 the distribution of the weights is localized around the value for an equal superposition of all 
 states in the $W = [000]$ winding sector, indicated by the vertically dashed line. The deviation from 
 this ideal result could become smaller with increase in the lattice size.

  Let us turn to the case of fermionic links, which is shown in the right panel of 
 \figref{state_histograms} in panels C and D, which plot the same physical quantities as those in A 
 and B, but for the fermionic model. For values of $\lambda$ up to the ground-state level crossing at 
 $\lambda_{c2} \approx 0.65$, the overall picture of the correlation histogram is very similar to the 
 bosonic case since the GS also is in the zero-winding sector. However, in contrast to the bosonic case, 
 many more states have zero weight (within machine precision). This could be due to some unresolved symmetry 
 for the fermionic model causing the coefficients of basis states with the same number of flippable 
 plaquettes to be equal and opposite, which then conspire to cancel out. A striking difference between the
 bosonic and the fermionic links occurs only above $\lambda_{c2}$, when the ground-state is in the $W=[002]$ 
 sector. We show the corresponding analysis of representative state at $\lambda = 0.8$ (green data) and 
 it immediately becomes apparent that most lattice configurations again have very small weights. Conversely, 
 the states with non-zero weight are observed to contribute equally to a very good approximation. This phase 
 therefore is indicative of an ordered phase which happens close to the RK point in the fermionic model. For 
 orientation, the green dashed-dotted line shows the corresponding amplitude of an equal superposition of all 
 states in the $W=[002]$ sector, and we see that all the contributing states are to the right of this line.

 The visual investigation of the structure of the GS wavefunction above, which suggests the delocalized 
nature of the wavefunction in the Hilbert-space in the potential spin-liquid phase, could also be made more 
concrete with other observables typically used to diagnose delocalization in Hilbert space \cite{Alet2018}. 
Specifically, we discuss the Shannon entropy, which can be written as a special case $S_1$ of the R\'enyi 
entropy of order $\alpha$:
\begin{equation}
  S_\alpha = \frac{1}{1-\alpha}\log\sum_{i=1}^N p_i^\alpha\ \ \stackrel{{\alpha\to 1}}{=}\ \ -\sum_{i=1}^N p_i\ln p_i,
\end{equation}
 where the probability $p_i = |\overlap{i}{\mathrm{GS}}|^2$ is the weight of the basis state $i$ in the 
ground state. Note that such entropies are dependent on the chosen many-body basis, but the values are not 
expected to be very different as long as the basis is not fine-tuned. The intuition behind this observable 
is a quantification of the amount of fluctuations in the ground-state wavefunction: While a maximally localized 
ground state would correspond to minimal values of $S_1$, the entropy grows with the amount of fluctuation to 
its maximal value when all states contribute equally (when it is maximally disordered).

\begin{figure}
  \includegraphics{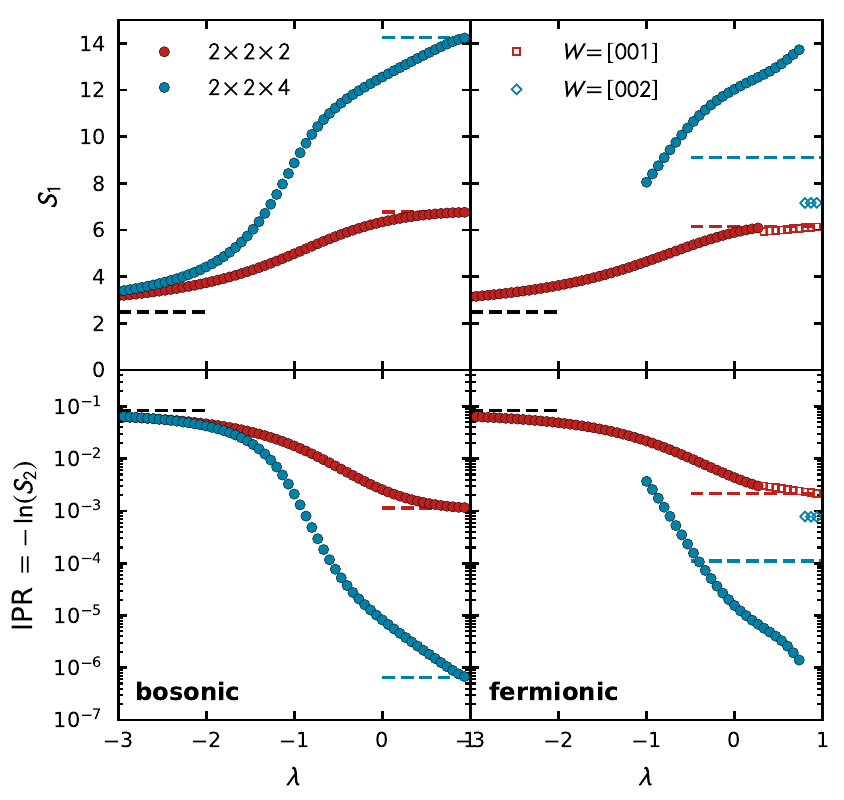}
  \caption{{\bf Entropy and IPR vs. $\lambda$} for bosonic (left) and fermionic (right) link variables are 
  shown in the top and bottom lines, respectively. The respective limits are shown with dashed lines. For 
  all plots, full symbols correspond to zero-winding states whereas open symbols (for fermions) correspond 
  to non-zero winding sectors, as indicated in the legend.}
  \label{fig:ipr}
\end{figure}

 In the top panels of \figref{ipr}, we show our numerical values of $S_1$ for bosonic (left) and fermionic 
(right) link models for two system sizes. In both cases, the ordered limit at $\lambda\to-\infty$ would 
correspond to $S_1 = \log(12)$, which is indicated by a black dashed line. We observe that $S_1$ for both 
system sizes converge to this limit. In the opposite limit, namely $\lambda\to 1$, we observe that the Shannon 
entropy for the bosonic link model quickly approaches the maximal value $S_1 = \log N_W$ where $N_W = |\mathcal{H}_W|$ 
denotes the size of the Hilbert space in the corresponding winding sector. This supports the picture from 
the above analysis, namely that the system exits the ordered phase as $\lambda$ is made small and positive, 
and enters a QSL phase which smoothly merges to the RK point $\lambda = 1$.

 For fermions a slightly different picture presents itself, which depends on the considered lattice size. 
While similar features to the bosonic case persist, above $\lambda_{c2}$ the entropy is expected to converge 
to $\log{N_{00z}}$, which is indicated by the colored dashed lines. This is indeed the case for the 
$\gsystem{2}{2}{2}$ lattice (for the $W_z = 1$ sector), where all states contribute almost equally irrespective 
of the $\lambda$ value. For the larger $\gsystem{2}{2}{4}$ lattice, $S_1$ settles at a smaller value, indicating 
that the ground state of the system now resides in a reduced number of states in the $W=[002]$ sector. This is 
completely consistent with the above analysis where the GS histograms in the same region showed that the weight 
of any state either vanishes or is approximately equal to all other non-zero weights.

For completeness, we also briefly discuss here a closely related quantity, namely the so-called inverse 
participation ratio (IPR) \cite{Alet2018}, defined via
\begin{equation}
  I = \sum_{i=1}^N {p_i^2}.
\end{equation}
  The IPR is an alternate probe for localization of a quantum state in a given many-body basis and is related 
to the R\'enyi entropy of order $\alpha = 2$ via $S_2 = -\log I$ such that both $I$ and $S_1$ encode similar 
information. We show our numerical values for the IPR in the lower panels of \figref{ipr}, where we can draw 
equivalent conclusions as for the entropy discussed above: While the ground-state wavefunction is dominated 
by the $12$ most flippable states at large negative $\lambda$ (convergence to the black dashed line at 
$I = 1/12$ irrespective of the system size) the IPR approaches the one of an equal superposition of all basis 
states in the given winding sector (appropriately colored dashed lines at $I=1/N$) and is, hence, reminiscent 
of a QSL-type behavior.

\subsection{Monopole string excitation}
\begin{figure}[b]
  \includegraphics{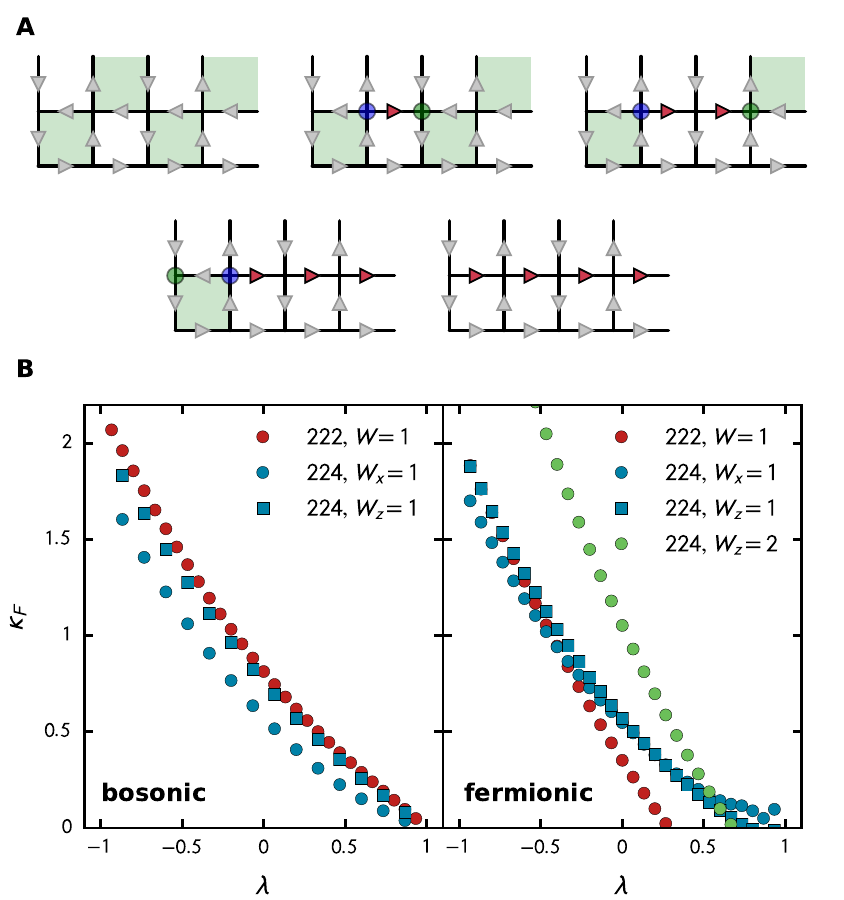}
  \caption{\label{fig:winding_gaps} {\bf Monopole string excitation.} (A) Sketch of the emergence of a 
  flux line. (B) $\kappa_F$ for two system sizes as a function of $\lambda$ for bosonic (left) and fermionic 
  (right) links. For the tube-like lattices, we plot two lines: one for fluxes along the short (disks) 
  direction and one for the long (squares) direction.}
\end{figure}

  One way to characterize the putative $U(1)$ liquid phase at $\lambda$ values close below the RK point 
is to investigate the cost of the flux lines in the system. The emergence of these flux-lines is illustrated 
in panel A of \figref{winding_gaps}: Starting from a state in the zero-winding sector (top left panel), 
where $G_x\ket{\psi} = 0$ on all vertices, we flip an arbitrary link $E_x,\mu$. Such a configuration is 
not in the pure-gauge sector, as Gauss' law at $x$ and $x+\hat{\mu}$ now corresponds to a positive and a 
negative charge sitting at these vertices, respectively. The flipped link acts as a ``flux line'' between 
these opposite charges (marked as the red arrows in the figure). Further, separating these charges prolongs 
these flux lines, until eventually the charges cross the boundary and annihilate each other leaving behind 
a line of flipped links with the condition $G_x\ket{\psi} = 0$ is again fulfilled at every vertex. As 
opposed to the initial basis state, the resulting state acquires a non-zero winding number and therefore 
does not belong to the zero-winding sector.

 In a confined phase, it is energetically costly to separate the charges, and the energy cost scales 
linearly with the distance between the pair of charges. The strength of this confinement is determined by 
the constant of proportionality, commonly referred to as the string tension $\sigma$, defined via 
$V(R) = \sigma R$ where $V(R)$ is the potential between two opposite charges. In a deconfined phase, as is 
expected in a $U(1)$ liquid phase, these ``flux excitations'' should cost much less energy, i.e., the string 
tension should be small. Of course, at finite system size this behavior is challenging to address precisely, 
however, the flux-excitations should be cheaper for $\lambda_c  \ge \lambda \ge -1$ (where we expect such 
a phase) than for the ordered state at large negative $\lambda$.

Following~\cite{Sikora2011}, we investigate the cost of such a flux-line excitation by measuring the 
monopole string tension, defined as
\begin{equation}
  \label{eq:mon_string_tension}
  \kappa_{\rm F} = \frac{E_{\rm F} - E_0}{L_{\rm F}} = \frac{\Delta_{\rm f}}{L_{\rm f}}
\end{equation}
where $E_{\rm F}$ is the ground state energy in the lowest non-zero flux sector, $E_0$ is the ground state 
in the zero-flux sector and $L_{\rm F}$ denotes the length of the flux tube (corresponding to the difference 
between the last and first panels of panel A in \figref{winding_gaps}, divided by the length of the flux line).

 Results for this quantity are shown in panel B of \figref{winding_gaps} for the two system sizes reachable 
with ED. For the bosonic case (left panel) we indeed observe the aforementioned trend, with a slight tendency 
to move towards the expected results for the thermodynamic limit (which is well beyond ED studies). There, 
$\kappa_{\rm F}$ should only be non-zero for $\lambda < \lambda_c$ but vanish above in the $U(1)$ liquid 
phase. Although the authors of Refs.~\cite{Sikora2009,Sikora2011} study the related dimer model (which has 
the identical Hamiltonian, but a different superselection sector) we find qualitatively similar results for 
the monopole string tension for the small systems studied here.

  For the fermionic case, on the other hand, the picture is slightly more complicated. There, the ground-state 
level crossing between the zero and non-zero winding sectors implies a vanishing cost of excitations for 
flux lines of arbitrary length. This is apparent in the right plot in panel B of \figref{winding_gaps} where 
$\kappa_F$ crosses over to negative values. We interpret this as a flux-condensed phase beginning, at least 
for the finite sizes considered here, already below the RK point. As the crossing point shifts to higher 
values of $\lambda$ with increasing system size, it will be interesting to see whether such a phase can be 
stabilized in the TL. While this implies that the RK point does not have its usual properties for the finite
size fermionic system, we expect the properties to be restored in the thermodynamic limit. In particular, we 
would expect the GS to have an equal superposition of basis states from all the different winding sectors for 
larger system sizes.


\section{Conclusions and Outlook}
 In this article, we have introduced the particle representation for the Abelian $U(1)$ QLMs, distinct from the
rishon representations considered before \cite{Brower1999}. We noted that while the physics of using hardcore
bosons (which are equivalent to using quantum spins) to represent the link operators give identical physics as 
that of using fermions in two-spatial dimensions, the physics is qualitatively different in three-spatial 
dimensions. Interpreting the plaquette term as a correlated hop of \emph{two} fermionic particles along two
adjacent lines diagonally intersecting the midpoint of perpendicularly-oriented links, as shown in \figref{plaq}, 
we could show that these particles have subdimensional motion under the kinetic term. For two spatial dimensions,
this implied a linear motion of the particles, and any negative signs due to fermions crossing the periodic
boundary gets cancelled by the hop of the paired fermion. In three-spatial dimensions, the fermions move along
a plane in two-spatial dimension and give rise to an opposite sign to a worldline as a boson. Therefore, the
fermionic operators give rise to different physics as the bosons in three and higher dimensions. 

 Using techniques of finite size scaling on results obtained from lattices up to 72-links, we showed that both
the fermionic and the bosonic model spontaneously broke the lattice rotation and the charge conjugation symmetry
for large and negative $\lambda$. On decreasing $\lambda$ towards zero, our results indicated that the symmetry
breaking gap dissolves and the gap does not decrease. We introduced the state-participation histogram as a tool
to identify any sign of symmetry breaking in the ground state wavefunction. This diagnostic clearly indicates
the absence of any special basis state in the region $\lambda \sim 0$ with any significant overlap with the
ground state wavefunction. In addition, a host of other observables such as the mass gap, the fidelity 
susceptibility, as well as several related information theoretic quantities reveal the absence of any symmetry
broken phase, and provides an indirect evidence of a $U(1)$ Coulomb phase, which is sometimes identified as
a spin-liquid phase in the condensed matter theory literature. Both the fermionic and the bosonic model show
these features, even though the potential liquid phase is possibly very different in the models. For example,
the winding strings are much easier to excite in the regions $\lambda \sim 0$ of the bosonic model, as compared 
to the fermionic model.

 Of course, the existence of this liquid phase is fascinating and needs to be investigated on larger lattices.
One could marginally extend our calculation by exploiting various commuting symmetries to reach larger system
sizes using exact diagonalization. However, this is likely going to be insufficient, and some stochastic
Monte-Carlo methods, or tensor network methods would be more useful to achieve significant progress here. 

 For the fermionic model, we observe a novel phase which is sandwiched between the liquid phase and the
RK point. This phase has condensed phases in the ground state, and is reminiscent of the staggered phase 
(which is stable for $\lambda > 1$ in both the models in two-spatial dimensions. However, the ground state
has winding strings only in the longer direction and none along the shorter direction. This could be a finite
size effect, disappearing in the thermodynamic limit to the usual RK point where the ground state is spread 
over all the winding sectors. 

  With the rapid advance in the field of quantum simulators using cold atoms or with Rydberg atoms, it is natural 
to imagine that such platforms can be used to realize the model proposed here. Such an experimental realization
would not only enable an independent verification of the physics proposed here, but also enable the study of
quantum dynamics in this model which could possibly be beyond the reach of any numerical method in the near
future. However, the three-spatial dimension involved in the problem provides a difficult challenge given
the well-known difficulty of implementing the plaquette interaction in a cold atom simulator \cite{Gonzalez2017}. 
However, we note that the possibility of realizing the plaquette interaction as a correlated hop could be
an altogether practical route to realize the four body interaction without the need of any auxiliary qubits
as in a digital simulation scheme \cite{Bender2018}. In this regard, our interpretation of the plaquette term as a 
correlated hop of particles could already extend the schemes provided in \cite{Ott2021}, which however was postulated
for the limit of large boson occupation numbers. In fact, Reference \cite{Guardado2021} already describes a 
Rydberg atom implementation of the fermionic $t$-$V$ model which allows the fermions to hop along only one 
spatial direction. With additional interactions to force fermions in adjacent chains to hop together, one 
could realize exactly the plaquette interaction. 

 Independently of the experimental realization, this class of models opens up some interesting avenues of
research purely from a quantum field theory perspective. One of the immediate question is to ask if the Coulomb
phase does indeed exist in the bosonic model in three-spatial dimension, is it possible to use dimensional 
reduction to obtain a confined continuum gauge theory in two-spatial dimension, which the so-called
D-theory approach advocates \cite{Wiese2021}. Other interesting questions include the formulation of the
field theory of a Coulomb phase in three-spatial dimensions, where the gauge fields are fermionic in nature,
the mechanism of including larger representations using fermionic states, as well as non-Abelian generalizations
of the fermionic links. The presence of possible critical points and the phase diagrams in such models, together
with synergies with experiments promise an exciting road ahead. 

\acknowledgements
We acknowledge fruitful discussions with Arnab Sen, Inti Sodemann, Joel~Steinegger, Shailesh Chandrasekharan and 
Uwe-Jens Wiese. L.R. is supported by FP7/ERC Consolidator Grant QSIMCORR, No. 771891, and the Deutsche 
Forschungsgemeinschaft (DFG, German Research Foundation) under Germany’s Excellence Strategy --EXC--2111--390814868. Research of E.H. at the Perimeter Institute is supported in part by the Government of Canada through the Department of Innovation, Science and Economic Development and by the Province of Ontario through the Ministry of Colleges and Universities.


\appendix
\section{\label{app:state_scaling} Constructing the GLS}
 The first step of the diagonalization routine consists of building the many-body Hilbert space of permissible 
states (i.e., those that fulfill the Gauss law on every vertex). A naive approach, i.e., listing all states 
and simply discard the ones that violate the constraint, is not useful because of the prohibitive scaling of 
the number of states. Here we offer two options how to efficiently construct the Hilbert space.

 \paragraph*{Recursive state search I} One way to construct all permissible states for bipartite lattices is 
to first divide the problem into the two sublattices. We call the sublattice $\Lambda_A$ the one where we 
place vertices that obey Gauss' law (see main text) and the sublattice $\Lambda_B$ which contains the 
``in-between'' vertices where we need to check for the validity of Gauss' law. Then, one can proceed with 
a recursive function as follows:
\begin{enumerate}
  \item The (recursive) function should take in a list as its argument.
  \item If the list is of the length of the sublattice $\Lambda_A$, return the current list. The recursion 
  is done and we have obtained a valid state.
  \item If the length of the list is shorter, then loop through all allowed vertices by Gauss' law:
  \begin{enumerate}
    \item Append the current vertex to the original list.
    \item With the updated list, check which vertices of $\Lambda_B$ are already surrounded by vertices on 
    $\Lambda_A$. If there are surrounded vertices in $\Lambda_B$, check if Gauss' law is satisfied on this 
    vertex. If not, terminate the recursion for this branch and return nothing - Gauss' law is violated and 
    hence the state is not valid. If Gauss' law is satisfied, or if there are not surrounded vertices in 
    $\Lambda_B$, call the recursion with the updated list.
  \end{enumerate}
\end{enumerate}

 Calling the function with an initially empty list creates a tree which is checked for validity on the fly. 
Only when the desired depth is reached (i.e., the size of the sublattice $A$) a state will be added to the
list in the end. Therefore, once the recursion is finished, only the valid GLS survive.

 Note that technically this could be problematic because for large systems this could lead to steep memory 
requirements. It is advisable to store the already obtained states to file at 
intermediate steps (buffered, for every $10^6$ states for instance), in order to keep the list short.

 \paragraph*{Recursive state search II} The allowed GLS can also be found by a nested application of 
the ``kinetic'' part of the Hamiltonian to a set of seed states (the plaquette flipping term). This is 
efficient, but needs at least one flippable basis state for each winding sector in order to be useful. 
Moreover, there are two potential caveats: 1) unflippable configurations cannot be reached (which likely 
is not an issue except at the RK point) and 2) this strategy relies on the ergodicity of the problem. 
It could be, for instance, that the Hilbert space does fragment into several subspaces due to some 
hidden symmetry. Then this strategy will in general not find all relevant states 
(see also~\cite{Sikora2011}).

\begin{figure}
  \includegraphics{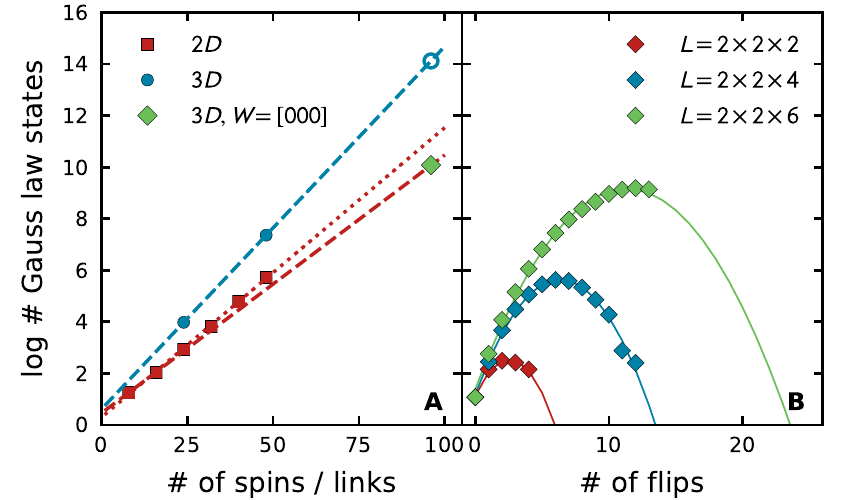}
  \caption{{\bf Scaling of the number of GLS.} (A) Total number of GLS for 2D (red) and 3D (blue) lattices,
  irrespective of winding. The dashed lines are an extrapolation with only the lowest data points, the 
  dotted line uses all available 2D data points. The green diamond is an estimate for the size of the 
  $W=[000]$ sector on a $\gsystem{2}{2}{6}$ lattice. (B) Number of GLS in the $W=[000]$ sector as 
  function of the flip level. Solid lines represent a quadratic fit.}
  \label{fig:state_number}
\end{figure}

  For completeness, we show the scaling of the total number of GLS for 2D and 3D lattices (across 
all winding sectors) in panel A of \figref{state_number}, where exact values are represented by filled 
symbols and (linear) extrapolations on the log scale are shown as dashed and dotted lines. Clearly, the 
requirements for a $\gsystem{2}{2}{6}$ lattice are steep, and it is likely that only the zero-winding
sector of this lattice size could potentially be reached with ED (green diamond) at, which is still of 
the order of $\sim 2^{33}$ states (without the consideration of further symmetries, e.g., translational 
invariance).

\section{\label{app:low_energy_details} Convergence of low-energy approximation}
 As briefly discussed in the main text, we employ a systematic low-energy approximation for the 
 largest considered systems in order to avoid the prohibitive scaling of the computational effort, 
 allowing us to gain some information on systems larger than $V = \gsystem{2}{2}{4}$. Here we 
 present some details of this approach.

\begin{figure*}[t]
  \includegraphics[scale=1.0]{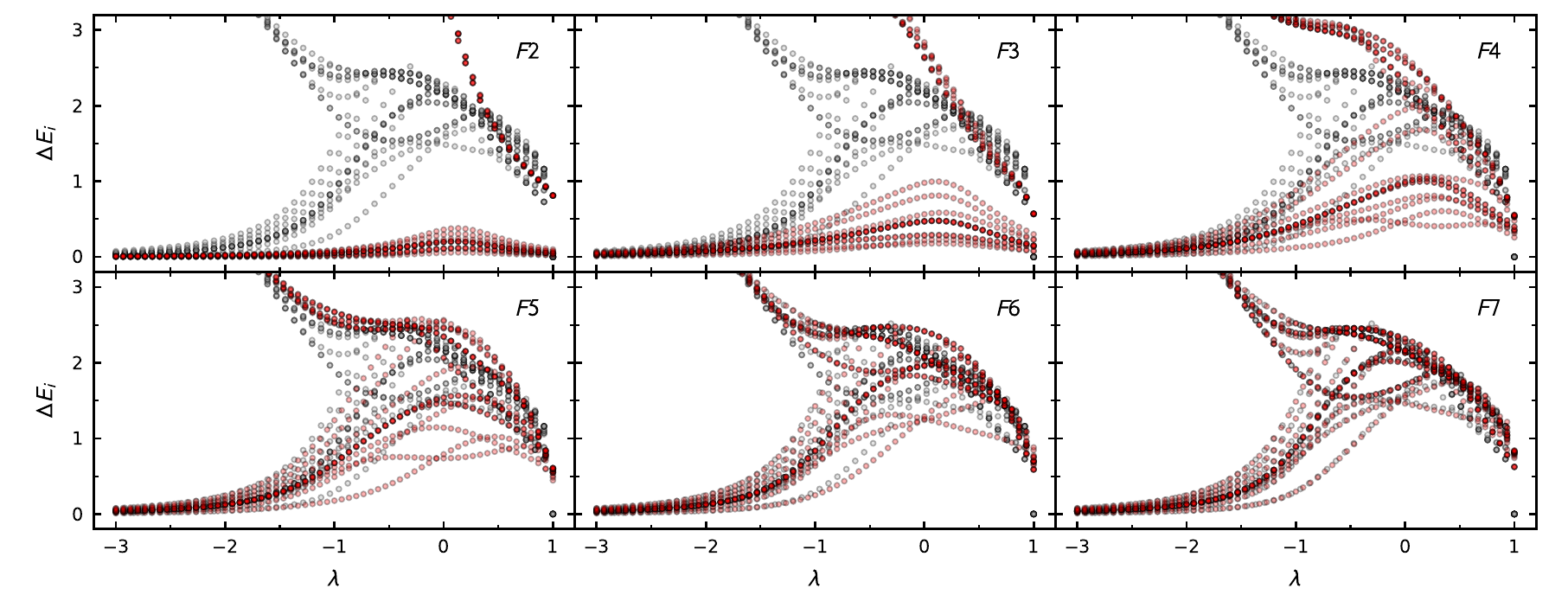}
  \caption{{\bf Low-energy spectrum of a bosonic $\gsystem{2}{2}{4}$ system} with all states in 
  the zero-flux sector considered (gray symbols) and with a restricted Hilbert space constructed 
  from plaquette flips (red symbols). From top left to bottom right: $2, 3, 4, 5, 6 $ and $7$ 
  flips away from the maximally flippable manifold, respectively.
  Darker coloring indicates degeneracy.}
  \label{fig:le_spectrum_convergence}
\end{figure*}

  The general idea is to first consider the ground state manifold in the limit $\lambda \to -
\infty$, which is a set of superposition of the most flippable lattice configurations. In this 
limit, the lowest $12$ states (this is the size of the GS manifold) could be extracted exactly 
by only considering the most flippable lattices. To systematically improve the obtained energies 
at $\lambda > -\infty$ we simply introduce ``excitations'' to the system by flipping single 
plaquettes. The extended set of states is expected to improve the approximation, as we now 
couple the low-lying states to excited states. Systematically repeating this sequence allows 
us to study the spectrum at different ``flip levels'' (FL) - this is shown in 
\figref{le_spectrum_convergence} for a $\gsystem{2}{2}{4}$ involving FL2 to FL7. At large 
negative $\lambda$, the spectrum converges quickly, since only small corrections are expected 
to be of importance in this regime. In the region $\lambda \approx 0$ convergence sets in only 
at lager FL, indicating the importance of states with an arbitrary number of flip excitations. 
Clearly, at $\lambda = 1.0$ due to the massive degeneracy of states with finite flux content 
at the RK point, higher flip levels will be needed to establish convergence of the spectra.

 The values for the excitation energies within such an approximation at $\lambda=-3.0$ and 
$\lambda=0.0$ are shown in panel B of \figref{spectra} and is observed to give a satisfactory 
convergence even at lattices with $72$ link variables. For quantities other than energy gaps, 
however, convergence was observed to be more challenging.

  Finally, we show the scaling of the number of GLS as a function of the flip level in panel 
B of \figref{state_number} for different lattice sizes, and we also explicitly give the number 
and fraction of the total Hilbert space for a $\gsystem{2}{2}{4}$ lattice in \tabref{nstates_224}.

\begin{table}
  \begin{tabular}{ccc}
    FL & number of states & \% \\
    \hline
    $0$ & 12      & $ 0.00077\%$\\
    $1$ & 396     & $ 0.0255 \%$\\
    $2$ & 5132    & $ 0.3307 \%$\\
    $3$ & 35660   & $ 2.298 \%$\\
    $4$ & 151864  & $ 9.785 \%$\\
    $5$ & 436088  & $ 28.1 \%$\\
    $6$ & 860664  & $ 55.45 \%$\\
    $7$ & 1245688 & $ 80.26 \%$\\
    \hline
    all & 1552024 & $100\%$
  \end{tabular}
  \caption{\label{tab:nstates_224}Number of states for the respective low-energy 
  approximations of the $\gsystem{2}{2}{4}$ lattice in the $[000]$ winding sector 
  compared to the full number of states (last line).}
\end{table}

\section{\label{app:diag} Diagonalizing the $2\times 2$ Fermionic and Bosonic Cases}
 In this Appendix, we explicitly work out the eigenvalues and the eigenvectors of $2 \times 2$ 
bosonic and fermionic systems, explicitly showing where they differ in negative signs. Before 
constructing the Hamiltonian explicitly for the $2 \times 2$ system, we can further simplify 
the analysis and the numerics by dividing the basis states into different winding number sectors
in $x$- and $y$-directions $(W_x, W_y)$. The winding number commutes with the Hamiltonian and 
in the electric flux basis, the Hamiltonian is block diagonal. Out of a total of 18 states, 
6 are in the zero winding sector.
The ground state lies in the zero winding sector, and hence is characterized by a six-dimensional 
vector. Using the action of the Hamiltonian given in (\ref{actions}), the eigenvalues are obtained 
by diagonalizing the matrix:
\begin{equation}\label{eq:bosH2x2}
   H_{(0,0)} = -J \left(\begin{array}{cccccc} 0&1&0&0&1&0\\
                    1&0&1&1&0&1\\
				    0&1&0&0&1&0\\
				    0&1&0&0&1&0\\
				    1&0&1&1&0&1\\
				    0&1&0&0&1&0 \end{array}\right).
\end{equation}
For J=1, the eigenvalues are  -2.82843, 0, 0, 0, 0, 2.82843.

 Before doing a similar exercise with the fermionic version, we need to label our states according 
to a definite convention, since fermionic operators are involved. The matrix which one obtains is
\begin{equation} \label{eq:fermH2x2}
   H_{(0,0)} = -J \left(\begin{array}{cccccc}
            0 & 1 & 0 & 0 & -1 & 0\\
            1 & 0 & 1 & -1 & 0 & -1\\
            0 & 1 & 0 & 0 & -1 & 0\\
	        0 & -1 & 0 & 0 & 1 & 0\\
	       -1 & 0 & -1 & 1 & 0 & 1\\
	        0 & -1 & 0 & 0 & 1 & 0
				    \end{array}\right)
\end{equation}
Indeed, on diagonalizing matrix (\ref{eq:fermH2x2}) we recover the same spectrum as that of 
(\ref{eq:bosH2x2}) for the quantum spin model.

\section{\label{app:fermi_exchange} Exchange of two fermions in 3D}
 In this appendix, we show the exact calculation demonstrating the emergence of a sign in the 
fermionic QLM, as argued in the main text (c.f. \figref{fermi_exchange}, we write down 
mathematically the flips that were considered in that figure). To do this, we first need to 
introduce the conventions regarding the ordering of operators, such that we are able to properly
identify an overall sign. In our conventions, \emph{normal ordering} refers to creation operators 
arranged so that their index is ascending from right to left, i.e., those with lower index are 
applied first.

\begin{figure}
  \includegraphics[width=0.8\columnwidth]{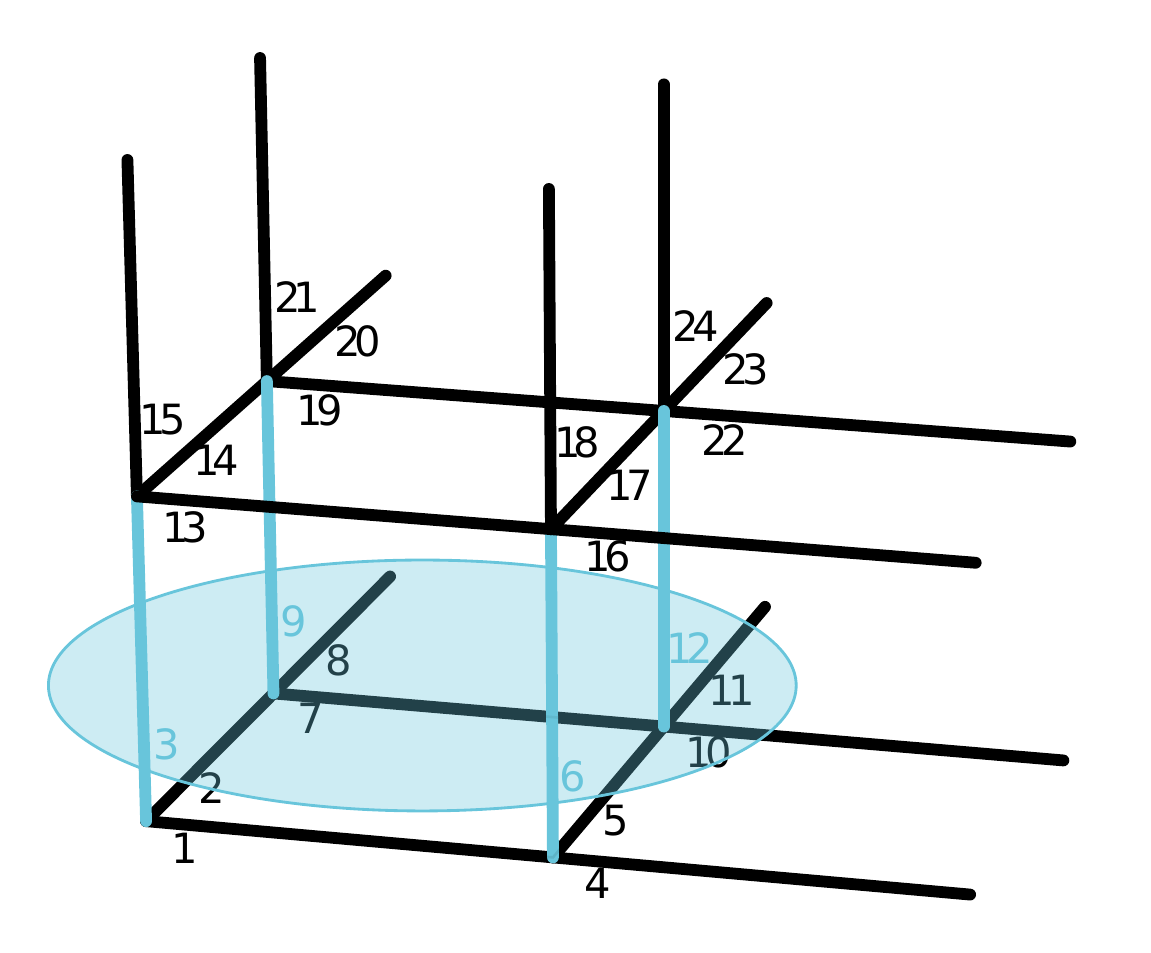}
  \caption{\label{fig:link_numbering}{\bf Numbering of links on a $\gsystem{2}{2}{2}$ lattice.} 
  The blue colored links $[3, 6, 9, 12]$ are summed to obtain the total flux $W_z$ in the 
  $z$-direction, indicated as the flux through a closed loop around the system in the $z$-plane. 
  Accordingly, $W_x$ is obtained by summing the links $[1, 7, 13, 19]$ and $W_y$ via summation 
  of the links $[2,5,14,17]$.}
\end{figure}

 Although the specific numbering of link variables merely is a technical detail, we show our 
convention in \figref{link_numbering} for the sake of completeness. In this convention, the 
initial state discussed in the main text corresponds to
\begin{equation}
  \ket{\psi_0} = \cre{24}\cre{12}\cre{11}\cre{10}\cre{7}\cre{5}\cre{4}\cre{1}\ket{0}.
\end{equation}

Moreover, we denote the plaquette operators as,
\begin{align}
  U_\square(i,j,k,l) &=  \cre{i}\cre{j}\ani{k}\ani{l},\\
  U^\dagger_\square(i,j,k,l) &= \ani{i}\ani{j}\cre{k}\cre{l},
\end{align}
where the indices denote the links of the plaquette in question.

  With this, we are able to explicitly show the emergence of a sign for the example discussed 
in the main text, specifically the string of plaquette operators applied in panel B of 
\figref{fermi_exchange}. For the sake of brevity, we only consider explicitly the application 
of the first of six plaquette operators, namely
\begin{align}
  \ket{\psi_1} &= U_\square^\dagger(5,12,17,6) \ket{\psi_0} \nonumber\\
  &= \left[\ani{5}\ani{12}\cre{17}\cre{6}\right] 
  \left[\cre{24}\cre{12}\cre{11}\cre{10}\cre{7}\cre{5}\cre{4}\cre{1}\ket{0}\right] \nonumber\\
  &= \cre{24}\cre{17}\cre{11}\cre{10}\cre{7}\cre{6}\cre{4}\cre{1}\ket{0},
\end{align}
where the last line is obtained by exploiting the usual fermionic anti-commutation relations. 
The subsequent operators generate the following sequence of states:
\begin{align}
  \ket{\psi_2} &= U_\square^\dagger(1,6,13,3) \ket{\psi_1}\nonumber\\
  &= -\cre{24}\cre{17}\cre{13}\cre{11}\cre{10}\cre{7}\cre{4}\cre{3}\ket{0} \\
  \ket{\psi_3} &= U_\square^\dagger(13,17,19,14) \ket{\psi_2}\nonumber\\
  &= -\cre{24}\cre{19}\cre{14}\cre{11}\cre{10}\cre{7}\cre{4}\cre{3}\ket{0} \\
  \ket{\psi_4} &= U_\square(2,9,14,3) \ket{\psi_3}\nonumber\\
  &= -\cre{24}\cre{19}\cre{11}\cre{10}\cre{9}\cre{7}\cre{4}\cre{2}\ket{0} \\
  \ket{\psi_5} &= U_\square(1,5,7,2) \ket{\psi_4}\nonumber\\
  &= -\cre{24}\cre{19}\cre{11}\cre{10}\cre{9}\cre{5}\cre{4}\cre{1}\ket{0} \\
  \ket{\psi_6} &= U_\square(7,12,19,9) \ket{\psi_5} \nonumber \\
  &= -\cre{24}\cre{12}\cre{11}\cre{10}\cre{7}\cre{5}\cre{4}\cre{1}\ket{0}
  = -\ket{\psi_0}.
\end{align}
As is evident from the last line, the application of these six operators (in this specific order) 
maps the state back to itself but with a sign that distinguishes a system with bosonic link 
variables from its fermionic counterpart.

\bibliography{refs_fqlm}
\end{document}